\documentclass[twocolumn,secnumarabic,assume, nobibnotes, aps, prd]{revtex4-2}

\usepackage{amsmath}
\usepackage{amsfonts}
\usepackage{amssymb}

\usepackage{psfrag,graphicx}
\usepackage{float}
\usepackage[T1]{fontenc}
\usepackage{xcolor}
\usepackage{lmodern}
\usepackage{listings}
\usepackage{subfig}
\usepackage[section]{placeins}
\usepackage{psfrag}
\usepackage{rotating}
\usepackage{multirow}
\usepackage{colortbl}
\allowdisplaybreaks
\usepackage{parskip}
\usepackage[T1]{fontenc}
\usepackage{siunitx} 

\usepackage{booktabs,caption}
\usepackage{threeparttable}
\newcommand{\blfootnote}[1]{{\renewcommand{\thefootnote}{\roman{footnote}}\footnotetext[0]{#1}}}
\begin{document} 
\title{Elastic and Inelastic Scattering of Flat-Top Solitons}

\author{U. Al Khawaja$^{1,*}$\blfootnote{*u.alkhawaja@ju.edu.jo}, M. O. D. Alotaibi$^{\,2}$, and L. Al Sakkaf$^{\,3}$\\\it{$^1$Department of Physics, School of Science, The University of Jordan, Amman, 11942, Jordan\\
		$^2$Department of Physics, Kuwait University, P.O. Box 5969 Safat, 13060 Kuwait\\
		$^3$Department of Physics, United Arab Emirates University, P.O. Box 15551, Al-Ain, United Arab Emirates}}
\begin{abstract}
We investigate the interaction between two flat-top solitons within the cubic-quintic nonlinear Schr\"odinger equation framework. Our study results point towards a significant departure of flat-top solitons collisional characteristics from the conventional behaviors exhibited in the scattering dynamics of two bright solitons. Our investigation outlines specific regimes corresponding to the dual flat-top solitons' elastic and inelastic collisions.  We determine regimes in the parameter space where exchange in the widths of the interacting flat-top solitons is possible, even within the elastic collision domain.  We find a periodic occurrence of completely elastic scattering of flat-top solitons in terms of the solitons' parameters. Investigating the internal energy transfers between the solitons and the emitted radiation reveals the origin of inelasticity in flat-top solitons collisions.  We perform a variational calculation that  accounts for the amount of radiation produced by the collision and hence provides further insight on the  physics underlying the loss of elasticity of collisions. 
\end{abstract}

\maketitle

\section{Introduction}
\label{sec:Introduction}
The most interesting and appealing feature of solitons is preserving their integrity while propagating and after scattering with each other \cite{books1,books2,books3,books4}.  Such a unique feature suggested solitons for a plethora of applications \cite{books5,books6,books7,books8}. Within this context, bright solitons of the fundamental nonlinear Schr\"odinger equation (NLSE) have  been extensively investigated \cite{books1,books2,books3}. Specifically, the interaction between bright solitons \cite{gordon} and their scattering by potentials \cite{kay,goodman,lee,ernst,usama3} are well studied. It is established that collisions between bright  solitons of the fundamental NLSE are elastic. It is understood that this is a consequence of the integrability of the fundamental NLSE which implies the existence of an infinite number of conserved  quantities (See for instance \cite{books1}.).

Collisions of solitons have been extensively studied across various models \cite{coll1,coll2,coll3,coll4,coll5}. The interactions of cosine-Hermite-Gaussian solitons have been analyzed using two-dimensional NLSE \cite{coll6}. Collisions between anisotropic dipolar solitons at different velocities and angles, examined through three-dimensional mean-field Gross-Pitaevskii equation, show quasi-elastic behavior at high velocities with no deformation and inelastic behavior with increasing deformation at lower velocities \cite{coll7}. Inelastic soliton collisions have also been explored in the nonlinear Klein-Gordon model \cite{coll8}. Additionally, the (3+1)-dimensional Sharma-Tasso-Olver-like model, applicable to dispersive wave phenomena in optics, reveals inelastic fussion or fission phenomena where some kink waves disappear or split into multiple waves \cite{coll9}.

Recently, an increased interest was directed towards another  type of solitons, known as flat-top solitons \cite{cheiney,bottcher,luo,9,10,11,12,13,14,15,16,17,18}. This was triggered by the experimental realization of the so-called quantum droplets, which are one form of flat-top solitons in  mixtures of Bose-Einstein condensates \cite{exp1,exp2,exp3,exp4} and dipolar condensates \cite{exp5,exp6,exp7}. Flat-top solitons are supported by the NLSE with dual power law nonlinearity. In mixtures of Bose-Einstein condensates, the competing nonlinearities are quartic and cubic. The former originates from quantum fluctuations, known as the Lee-Huang-Yang (LHY)  fluctuations \cite{LHY1,LHY2}, while the latter originates from the interatomic interactions \cite{hcbook}. In nonlinear optics, dual nonlinearities are cubic and quintic, where the former, known as the Kerr nonlinearity, corresponds to the nonlinear response of the medium, and the latter corresponds to higher order corrections \cite{books6}.  

Many features of flat-top solitons have been studied including their dynamics \cite{bor,25,26,27,28}, collective modes \cite{24}, their scattering by potentials \cite{33}, and binding into molecules \cite{usnew}. It is known that the NLSE with dual nonlinearities is not integrable \cite{books1}. Collisions of  flat-top solitons are also known to be inelastic  \cite{abd} and lead to fragmentation of solitons \cite{bor}. Given the importance of understanding the behavior of flat-top solitons from fundamental and applications point of views, we aim in this paper at investigating mutual collisions of flat-top solitons with emphasis on elasticity and inelasticity of the collisions. In particular, we will answer the following questions: 1) May collisions of flat-top solitons be elastic?  2) What is the physics underlying inelasticity in the collisions of flat-top solitons? To answer these questions, we start by characterizing the outcome of collisions between flat-top solitons in terms of the solitons' parameters, namely initial relative phase, initial relative speed, initial separation, and soliton width, where the latter being determined by the strength of the  nonlinearities. This results in a classification of scattering regimes into: 1) nearly elastic scattering associated with width exchange, 2) inelastic scattering with radiation production, and 3) a single case of completely elastic scattering with no width exchange and no emitted radiation. Investigating the single case of completely elastic scattering, we find periodic occurrences of such a case in terms of any  of the initial parameters of flat-top solitons, including their relative phase, separation, and speed. The second main finding is revealing the origin of inelasticity through a detailed analysis of the energy transfers between the different forms of energy. Specifically, we consider a completely inelastic case where the two solitons coalesce after collision, and then show that the permanent loss in the initial center-of-mass kinetic energy is mostly carried by dispersive radiation waves. We discuss this mechanism with respect to the nonintegrability of the NLSE with dual nonlinearity. This picture is confirmed through a variational calculation where we treat  the amplitude of the produced radiation as a variational parameter. Conservation of norm, energy, and momentum are then used to predict the norm of radiation which agrees favorably with the numerical results.

The rest of the paper is organized as follows. In Section~\ref{modelandsetup}, we present the theoretical model and setup for the numerical simulations. In Section~\ref{compt}, we present and discuss our results for characterizing the outcome of flat-top solitons scattering in terms of solitons' parameters. In Section~\ref{sub32}, we show results for periodic occurrences of elastic scattering in terms of the solitons parameters.   In Section~\ref{sub33}, we discuss energy transfers shedding more light on the origin of inelasticity.  In Section~\ref{analsec}, we perform a variational calculation to account for the amount of radiation production. We end in Section~\ref{conc} with a summary of our main findings, conclusions, and outlook for future work.

\newpage
\section{Theoretical Model and numerical setup}
\label{modelandsetup}
\subsection{Theoretical Model}
\label{sec:Theoretical_Model}
The dimensionless nonlinear Schr\"odinger  equation with dual power nonlinearities is expressed as
\begin{eqnarray}
	\label{dualnlse}
	i \frac{\partial}{\partial t} \psi(x, t)  &+& g_{1} \frac{\partial^2}{\partial x^2} \psi(x, t) + g_{2} |\psi(x, t)|^{p} \psi(x, t)\nonumber\\&+& g_{3} |\psi(x, t)|^{q} \psi(x, t) = 0,
\end{eqnarray}
where \( \psi(x, t) \) represents a complex field  corresponding, for example, to the amplitude of the electric (or magnetic) field, in case of optical solitons, while it corresponds to the condensate wavefunction, in case of Bose-Einstein condensates. The term \( g_{1} \) is the dispersion coefficient, while \( g_{2} \) and \( g_{3} \) denote the strengths of the nonlinearity terms, and $p$ and $q$ are arbitrary real constants that are not necessarily integers. This is one of the nonlinear evolution equations that admit exact solutions corresponding to flat-top solitons \cite{biswas}. Experimentally-relevant situations, such as quantum droplets of Bose-Einstein condensates, or optical solitons in fibers require $q=p/2$. In this case, Eq.~(\ref{dualnlse}) admits the following exact solution that corresponds to a one-parameter family comprising four fundamentally different solutions including flat-top solitons
\begin{eqnarray}
&&\psi(x,t)=\nonumber\\&&\left\{\frac{(p+2)u_0}{g_2+g_2\sqrt{1+\gamma}\,{\rm cosh}\left[p\sqrt{\frac{u_0}{g_1}}
(x-x_0-v_0t)\right]}\right\}^{1/p}\nonumber\\&\times&e^{i\left[u_0t+\frac{v_0}{4g_1}(2(x-x_0)-v_0t+\phi_0)\right]}
\label{ftgeneral}.
\end{eqnarray} 
In this equation, \( u_{0} \), \( x_{0} \), \( \phi_{0} \), and \( v_{0} \) are parameters that denote the soliton's initial amplitude, peak position, phase, and velocity, respectively. The dimensionless parameter \( \gamma \), expressed as \( \gamma= {g_{3}}/{g_{30}} \) where \( g_{30} ={(p+1) g_{2}^2}/({(p+2)^2 u_{0}}) \), is the single parameter that dictates the form of the soliton solution. Depending on \( \gamma \), the soliton may manifest as a  kink  soliton ($\gamma=-1$), flat-top soliton ($-1<\gamma<0$), bright soliton ($\gamma=0$), or thin-top soliton ($\gamma>0$), as discussed in \cite{Sakkaf1b}. 

\begin{figure}[!h]    
\includegraphics[width=\columnwidth]{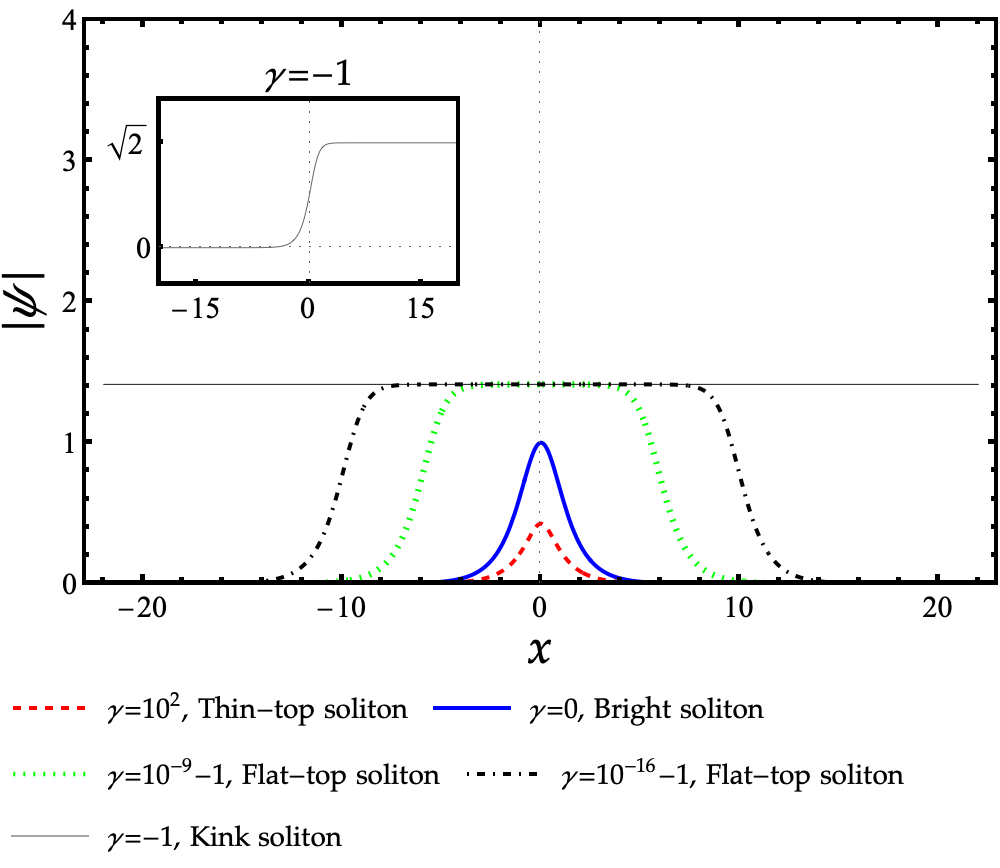}
\caption{ Solution profile of Eq.~\eqref{ftsol} for different values of $\gamma$. Values of other parameters are: $g_1=1/2,\,g_2=1,\,u_0=1$, $v_0=0$. } 
\label{fig1}
\end{figure}

For optical solitons in fibers, the nonlinearities are cubic and quintic, which correspond to $p=2$, and for this case, Eq.~(\ref{dualnlse}) takes the form
\begin{eqnarray}
	\label{cqnlse}
	i \frac{\partial}{\partial t} \psi(x, t)  &+& g_{1} \frac{\partial^2}{\partial x^2} \psi(x, t) + g_{2} |\psi(x, t)|^{2} \psi(x, t) \nonumber\\ &+& g_{3} |\psi(x, t)|^{4} \psi(x, t) = 0,
\end{eqnarray}
and its solution, (\ref{ftgeneral}), simplifies to
\begin{eqnarray}
	\label{ftsol}
\psi\left(x,t\right) &=& \sqrt{\frac{2u_{0}}{g_{2}\sqrt{1+\gamma}}} \nonumber\\&\times&
\frac{1}{\sqrt{\frac{1-\sqrt{1+\gamma}}{2\sqrt{1+\gamma}}+\mathrm{cosh}^2\left[\sqrt{\frac{u_{0}}{g_{1}}}(x-x_{0}-v_{0}t)\right]}}\nonumber\\&\times&  e^{i \left[u_{0}t + {v_{0}\left(2(x - x_{0}) - v_{0}t\right)}/({4 g_{1}})\right]}.
\end{eqnarray}
Figure~\ref{fig1} illustrates the soliton profiles corresponding to the four members of this   \( \gamma{\rm -family } \) of solutions.  

The width of the flat-top soliton, $W$, is defined by the half-maximum criterion, namely $|\psi(W/2,0)|=(1/2)|\psi(x_0,0)|$, which gives
\begin{equation}
W=2\sqrt{\frac{g_1}{u_0}}\ln{\left[4+\frac{3}{\sqrt{1+\gamma}}+\frac{1}{2}\,\sqrt{-4+\left(8+\frac{6}{\sqrt{1+\gamma}}\right)^2}\right]}
\label{width}.
\end{equation}

In the limit $\gamma\rightarrow-1$, where $\gamma$ approaches $-1$ from the right, the soliton width diverges, and the left and right edges of the flat-top soliton approach $-\infty$ and $+\infty$, respectively. The soliton, in this limit, will take the form  of a finite constant value, as shown by the thin horizontal line  in Fig.~\ref{fig1}. However, since $x_0$ is an arbitrary constant, the solution can be shifted by $x_0=W/2$ such that the left edge of the flat-top soliton is always at the origin. With this shift, the solution, in the limit $\gamma\rightarrow-1$, takes the shape of a kink localized at $x=0$, while the right edge is shifted to $+\infty$, as shown by the inset in Fig.~\ref{fig1}.  

%

\begin{figure}[!ht]    
\includegraphics[width=\columnwidth]{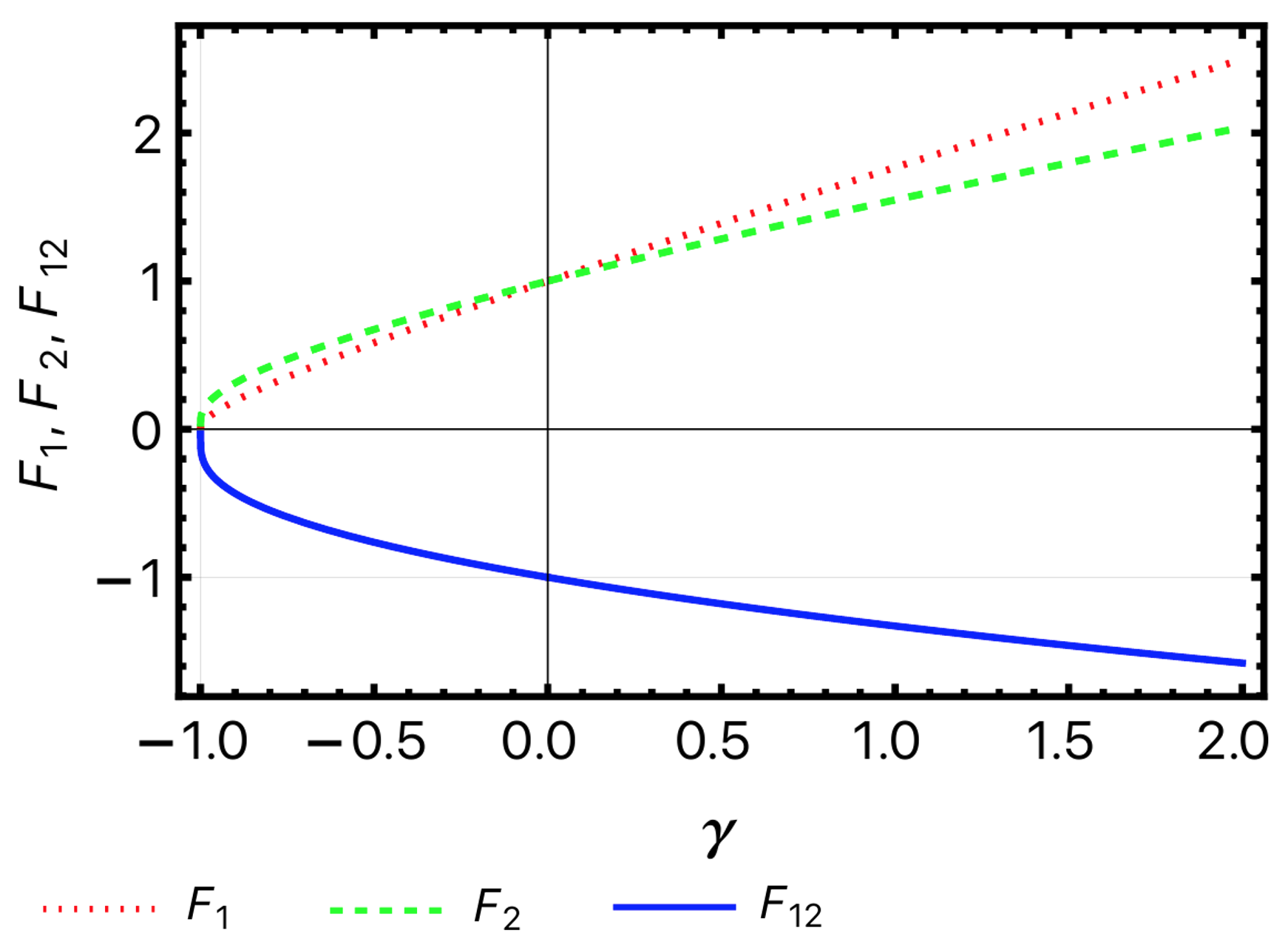}
\caption{The function $F_{12}=F_1-2F_2$ given by Eq.~(\ref{f12}). } 
\label{fig2}
\end{figure}

\begin{figure}[!ht]    
\includegraphics[width=\columnwidth]{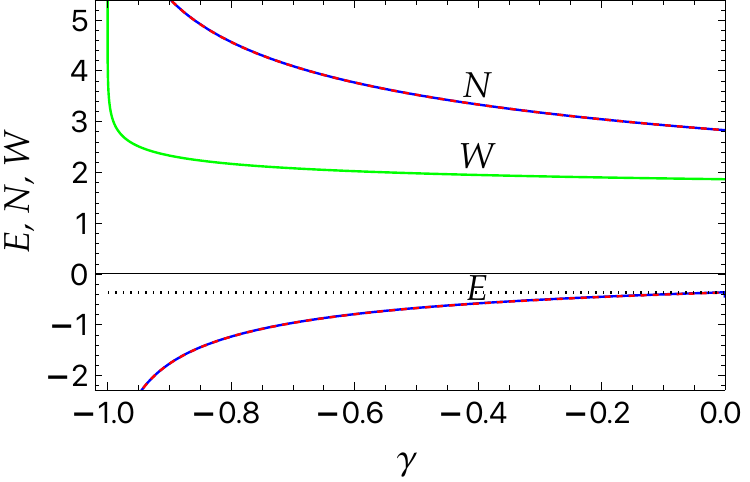}
\caption{Solid curves correspond to the analytical expressions of the total energy, $E$, norm, $N$, of two solitons and width, $W$, of a single soliton, as is essentially given by Eqs.~(\ref{etot3}), (\ref{norm}), and (\ref{width}), respectively. It should be noted that expression (\ref{etot3}) gives the energy of a single soliton. The corresponding curve in this figure is for two solitons; initially one is stationary and the other is moving with a speed $v_0$. Therefore, we have multiplied the interaction energy term in (\ref{etot3}) by 2. Dashed red curves for $E$ and $N$, which coincide with the analytical expressions, are obtained form the numerical solution. Horizontal dotted line marks the bright solitons energy limit. Parameters used: $u_0=1$, $g_1=1/2$, $g_2=1$.} 
\label{fig3}
\end{figure}

The energy functional corresponding to Eq.~(\ref{cqnlse}) is
\begin{equation}
E=\int_{-\infty}^{\infty}\left[g_1\left|\frac{\partial\psi}{\partial x}\right|^2-\frac{1}{2}g_2|\psi|^4-\frac{1}{3}g_3|\psi|^6\right]dx
\label{efunc}.
\end{equation}
The first term in the energy functional gives the kinetic energy which, for a quantum state, can be split into two terms, namely  the {\it center-of-mass kinetic energy} and the {\it kinetic energy pressure}. The former corresponds to the  translational motion of the flat-top soliton, which is analogous to the classical translational kinetic energy of rigid objects, while the second term originates from the curvature in the profile  and has no classical analog. The two terms can be extracted from the total kinetic energy term, $KE_{tot}=g_1\int_{-\infty}^{\infty}\left|\frac{\partial\psi}{\partial x}\right|^2dx$, as follows. Consider the polar form of the wavefunction: $\psi(x,t)=A(x,t)e^{i\phi(x,t)}$, with real amplitude and phase. The total kinetic energy term can then be written as 
\begin{equation}
{KE}_{tot}=g_1\int_{-\infty}^{\infty}\left(\frac{\partial A}{\partial x}\right)^2dx
+g_1\int_{-\infty}^{\infty}A^2\left(\frac{\partial \phi}{\partial x}\right)^2dx
\label{ketot},
\end{equation}
which defines
the kinetic energy pressure
\begin{equation}
KE_{pressure}=g_1\int_{-\infty}^{\infty}\left(\frac{\partial A}{\partial x}\right)^2dx\label{kepr}
\end{equation}
and center-of-mass kinetic energy
\begin{equation}
KE_{com}=g_1\int_{-\infty}^{\infty}A^2\left(\frac{\partial \phi}{\partial x}\right)^2dx\label{kecom}.
\end{equation}
It should be noted here that this distinction between the two forms of energy may not be well defined for the general case when amplitude, $A(x,t)$, and phase, $\phi(x,t)$, have general time dependence, since in that case, each of the two types of energy may have contributions from both definitions (\ref{kepr}) and (\ref{kecom}). However, for solutions which depend on time only linearly through their translational motion, obtained by a Galilean transformation of a stationary solution, such as the one given by (\ref{ftsol}), the two types of kinetic energy are well defined. In this case, using the solution (\ref{ftsol}), the kinetic energy pressure takes the form
\begin{equation}
KE_{pressure}=\frac{1}{48g_1}g_2^2N^3F_1(\gamma)
\label{kepr2},
\end{equation}
where 
\begin{eqnarray}
&&F_1(\gamma)=\frac{3}{16}\,{(-\gamma)^{-3/2}}\,{\left[\frac{\gamma}{{\rm tan}^{-1}\left(\frac{\sqrt{1+\gamma}-1}{\sqrt{\gamma}}\right)}\right]^{3/2}}\nonumber\\&\times&\left[\sqrt{-\gamma}-2(1+\gamma){\rm tanh}^{-1}\left(\sqrt{-\frac{2+\gamma-2\sqrt{1+\gamma}}{\gamma}}\right)\right]\nonumber\\
\end{eqnarray}
and the center-of-mass kinetic energy takes the form
\begin{equation}
KE_{com}=\frac{1}{4g_1}Nv_0^2
\label{kecom2},
\end{equation}
where $N$ is the total soliton intensity, or norm, defined by
\begin{eqnarray}
N&=&\int_{-\infty}^{\infty}|\psi|^2dx\nonumber\\\nonumber\\
&=&8\sqrt{\frac{g_1u_0}{g_2^2\gamma}}\tan^{-1}\left(\frac{\sqrt{1+\gamma}-1}{\sqrt{\gamma}}\right)
\label{norm}.
\end{eqnarray}
The total interaction energy is also calculated as
\begin{eqnarray}
IE&=&\int_{-\infty}^{\infty}\left[-\frac{1}{2}g_2|\psi|^4-\frac{1}{3}g_3|\psi|^6\right]dx\nonumber\\
&=&-\frac{1}{24g_1}g_2^2N^3F_2(\gamma)
\label{ie},
\end{eqnarray}\\
\begin{widetext}
where
\begin{eqnarray}
F_2(\gamma)&=&-\frac{3}{512 c^{5/2} (c+1)^{5/2}}\left\{
\frac{16 c (c+1)} {\gamma +1}\left[\frac{\gamma }{\tan ^{-1}\left(\frac{\sqrt{\gamma
   +1}-1}{\sqrt{\gamma }}\right)^2}\right]^{3/2} \left[\sqrt{c} \sqrt{c+1}-(2 c+1) \tanh
   ^{-1}\left(\sqrt{\frac{c}{c+1}}\right)\right]
\right.\nonumber\\&+&\left.
\gamma  \left[\frac{\gamma }{(\gamma +1) \tan ^{-1}\left(\frac{\sqrt{\gamma
   +1}-1}{\sqrt{\gamma }}\right)^2}\right]^{3/2} \left(3 \sqrt{c} \sqrt{c+1} (2
   c+1)-\left(8 c^2+8 c+3\right) \tanh ^{-1}\left(\sqrt{\frac{c}{c+1}}\right)\right)
   \right\}
\end{eqnarray}
\end{widetext}
and $c=\frac{1-\sqrt{\gamma +1}}{2 \sqrt{\gamma +1}}$. 
The total energy is thus
\begin{equation}
E=\frac{1}{4g_1}Nv_0^2+\frac{1}{48g_1}g_2^2N^3F_1(\gamma)-\frac{1}{24g_1}g_2^2N^3F_2(\gamma)
\label{etot2}.
\end{equation}
While the analysis we perform below on the flat-top soliton dynamics requires the determination of each energy component separetly, it is instructive to show that the last expression simplifies to 
\begin{equation}
E=\frac{1}{4g_1}Nv_0^2+\frac{1}{48g_1}g_2^2N^3\,F_{12}(\gamma)
\label{etot3},
\end{equation}
where
\begin{eqnarray}
F_{12}(\gamma)&=&F_1-2F_2=-\frac{3 }{4 \tanh ^{-1}\left(\frac{\sqrt{\gamma
   +1}-1}{\sqrt{-\gamma }}\right)^2}\nonumber\\&\times&\left[1+
   \frac{\sqrt{-\gamma }}{2 \tanh ^{-1}\left(\frac{\sqrt{\gamma
   +1}-1}{\sqrt{-\gamma }}\right)}\right]
\label{f12}.
\end{eqnarray}
The quantity $F_{12}(\gamma)$ is real for our relevant range of $\gamma$, namely $\gamma>-1$. Plotting this quantity, as shown in Fig. \ref{fig2}, shows that it equals $-1$ for the bright soliton ($\gamma=0$) and equals 0 for the kink soliton ($\gamma=-1$).  In the bright soliton limit, $\gamma\rightarrow0$, both functions $F_1(\gamma)$ and $F_2(\gamma)$ equal 1, and the energy takes the correct form of bright soliton energy, namely $E_{BS}=\frac{1}{4g_1}Nv_0^2-\frac{1}{48g_1}g_2^2N^3$. In the limit $\gamma\rightarrow-1$, which corresponds to the kink soliton, $N$ diverges as $-\ln{(1+\gamma)}$ while the term $F_{1}-2F_2$ vanishes as $1/(\ln{(1+\gamma)})^2$. Equation~(\ref{etot3}), shows that the energy, which is proportional to $N^3(F_1-2F_2)$, diverges as $-\ln{(1+\gamma)}$. Divergence in energy of the kink soliton is due to  the infinite nonzero flat part of the kink soliton, which results in a diverging norm. In Fig.~\ref{fig3}, we plot the width, $W$, the norm, $N$, and the energy, $E$, of a single moving soliton, as given by expressions (\ref{width}), (\ref{norm}), and (\ref{etot3}), respectively.

\subsection{Setup for Numerical Simulations}
\label{setupsec}
We consider two equal flat-top solitons; a stationary soliton located at $x=-x_0$ and a second soliton at $x=x_0>0$ moving towards the first soliton with initial center-of-mass velocity $v_0<0$. The profile corresponding to this initial setting is
\begin{equation}
	\psi\left(x,0\right) =\psi_l\left(x,0\right)+ \psi_r\left(x,0\right)
	\label{ftinit},
\end{equation}
where the initial profile of the stationary soliton is
\begin{equation}
	\label{ftinit0}
	\psi_l\left(x,0\right) =
 \sqrt{\frac{2u_{0}}{g_{2}\sqrt{1+\gamma}} }
\frac{1}{\sqrt{\frac{1-\sqrt{1+\gamma}}{2\sqrt{1+\gamma}}+\mathrm{cosh}^2\left[\sqrt{\frac{u_{0}}{g_{1}}}(x+x_0)\right]}},
\end{equation}
and the initial profile of the moving soliton is 
\begin{eqnarray}
\psi_r\left(x,0\right)& =& \sqrt{\frac{2u_{0}}{g_{2}\sqrt{1+\gamma}}} 
\frac{1}{\sqrt{\frac{1-\sqrt{1+\gamma}}{2\sqrt{1+\gamma}}+\mathrm{cosh}^2\left[\sqrt{\frac{u_{0}}{g_{1}}}(x-x_{0})\right]}}\nonumber\\&\times&  e^{i \left[ {v_{0}\left(x - x_{0} \right)}/({2 g_{1}})\right]},\nonumber\\
\label{ftr}
\end{eqnarray}
where  we have set the initial phase to  $\phi_0=0$, which will be the  case for most of the numerical experiments. In Section~\ref{sub32}, however, we relax this setting and allow this phase to change in order to investigate its effect on the periodic occurrence of elastic scattering.
This initial profile is evolved using Eq.~(\ref{cqnlse}). We use the iterative power series finite difference method \cite{ps1,ps2}, which proved to be highly accurate in simulating the scattering of very wide flat-top solitons by potential wells \cite{Sakkaf1b}. The method is superior over other finite difference methods in accuracy and speed. We have shown in Ref.~\cite{Sakkaf1b}  that accuracy can reach machine precision and speed is at least 3 times faster than split-step finite different method.

It is  instructive to find how the soliton's main features, namely energy, norm, and width, depend on the the parameter $\gamma$. These three quantities are plotted in Fig.~\ref{fig3}, for a single soliton moving with speed $v_0$, using  the exact analytical expressions (\ref{width}), (\ref{norm}), and (\ref{etot3}). As a confirmation on the accuracy of the numerical code, the energy and norm are also calculated numerically using the integrations in (\ref{kepr}), (\ref{kecom}), (\ref{norm}), and (\ref{ie}), where a perfect agreement with the analytical expressions is obtained.   

A final remark about the values of $\gamma$ that we consider in this work is in order. Our numerical investigations are restricted to flat-top solitons with $-1<\gamma<0$. For wide flat-top solitons, $\gamma$ has to be very close to $-1$. It is thus convenient to consider values of $\gamma$ given by the function: $\gamma_i=10^{-i}-1$, where $i=0,1,2,\dots,16$. In this manner, the width of the flat-top soliton, given by Eq.~(\ref{width}), increases almost linearly with $i$ for $i>2$. Here, $\gamma_0$ corresponds to the bright soliton and $\gamma_{16}$ corresponds to the widest flat-top soliton that we consider  here, with width $W\approx14.3$ for $g_1=1/2,\,u_0=1$.

\section{Computational Results and Analysis}
\label{compt}
Elastic scattering of classical rigid objects is characterized by the conservation of total kinetic energy. Numerical inspection showed that scattering of flat-top solitons is complicated by the various possible outcomes. Depending on the solitons initial relative speed, phase, and width, the scattering may be, as in the classical case, completely elastic, completely inelastic (coalescence), or partially inelastic. These possible outcomes are complicated by the possibility of permanent average {\it width exchange}, excitation of internal modes such as width oscillations, and emitted radiation. We found that, while the collision may be nearly elastic, considerable width exchange may take place  accompanied by very small width oscillations. It should be noted however that only for the completely elastic case, there is no width exchange, no width oscillations, no radiation, and the kinetic energy is exactly conserved. Emitted radiation is always present in flat-top scattering, except for the completely elastic scattering case, and turns out to provide the key for understanding the source of inelasticity in flat-top scattering.

In this section, we examine the dynamics of interactions between two flat-top solitons with emphasis on kinetic energy and width changes before and after scattering. In Sec.~\ref{sec:FTS_gamma9}, we select a flat-top soliton with a width corresponding to \(\gamma_{9}\) for the numerical study, outlining the parameters space for elastic and inelastic collisions, in addition to the width exchange between the two flat-top solitons post-collision. We also demonstrate that the two flat-top solitons merge into a single entity at very low and very high collision velocities, where the coalescence in both instances is accompanied by the ejection of  small-amplitude propagating radiation waves. In Sec.~\ref{sec:FTS_gamma0To16}, we broaden the study to encompass other flat-top solitons with varying widths corresponding to the range \(\gamma_{0}-\gamma_{16}\).

\subsection{Collision of Two Flat-Top Solitons with Widths Corresponding to \(\gamma_{9}\)}
\label{sec:FTS_gamma9}

\begin{figure*}[htbp]
  \includegraphics[width=1\textwidth]{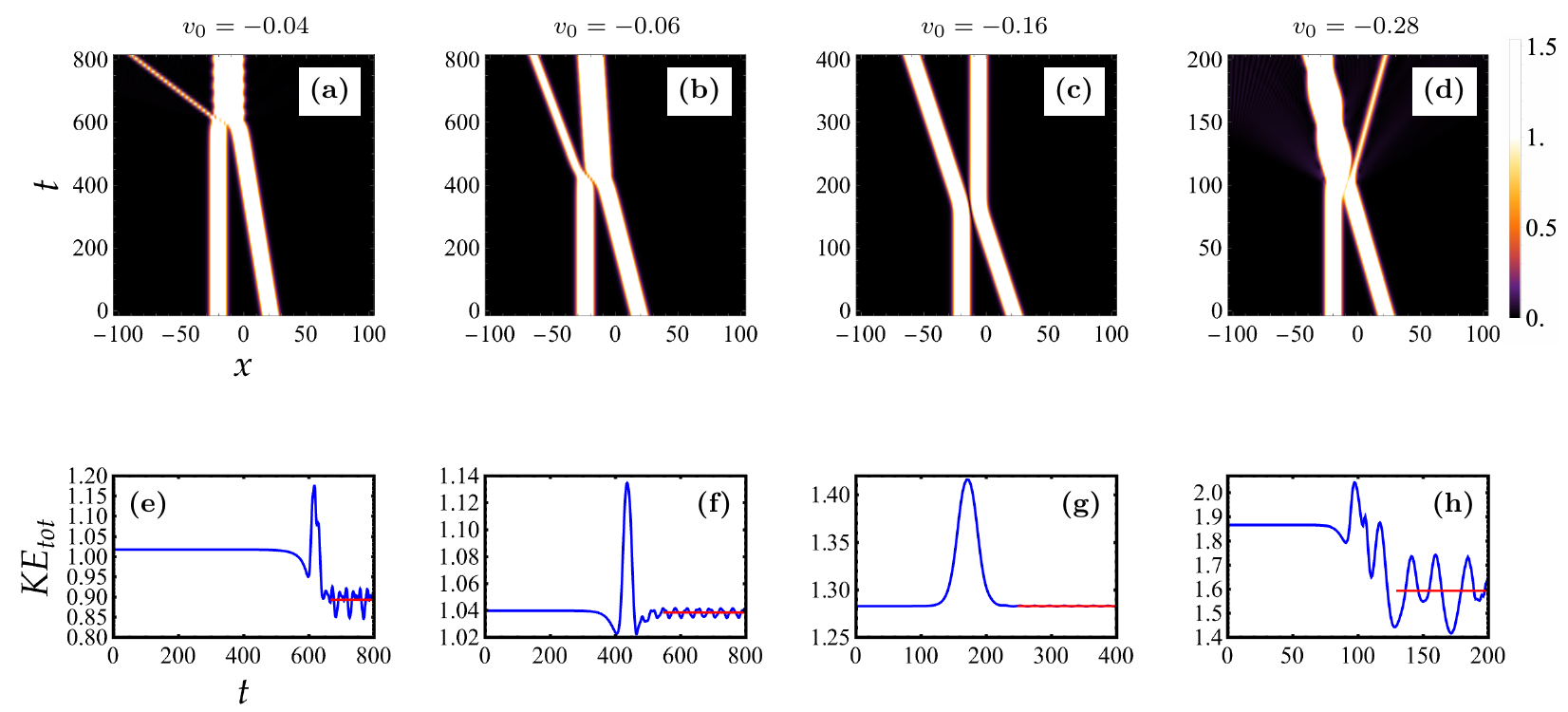}
\caption{Interaction of two flat-top solitons with widths defined by \( \gamma_{9} \) (upper row) and the corresponding  total kinetic energy as a function of time (lower row). The red line indicates the mean value. In each case, one soliton remains stationary while the other is moving towards it with a speed $v_0$. The velocity and  change in kinetic energy, respectively, are (a), (e) $v_0=-0.04$, $|\Delta \text{KE}|=1.243\times10^{-1}$, (b), (f) $v_0=-0.06$, $|\Delta \text{KE}|=1.374\times10^{-2}$, (c), (g) $v_0=-0.16$, $|\Delta \text{KE}|=2.074\times10^{-7}$, and (d), (h) $v_0=-0.28$, $|\Delta \text{KE}|=2.729\times10^{-1}$. Values of other parameters are: $g_1=1/2,\,g_2=1,\,u_0=1$. }
  \label{fig4}
\end{figure*}

For this part of our study, a flat-top soliton with a width corresponding to \( \gamma_{9} \) is used to highlight multiple outcomes. Total kinetic energy is calculated before and after collision using Eq.~(\ref{ketot}). The difference in kinetic energy is then defined as $\Delta KE=KE(t=t_f)-KE(t=0)$.
Figure~\ref{fig4} demonstrates the degree of elasticity observed in the collision between two flat-top solitons. In (a), a stationary flat-top soliton with velocity \( v_0 = 0 \) is approached by another flat-top soliton with \( v_0= -0.04 \). The collision is inelastic, as evidenced by the difference in kinetic energy before the collision and the average kinetic energy afterward, represented by the red line, accompanied  by a significant width exchange. In (b), the flat-top soliton's velocity is increased to \( v_0= -0.06 \), resulting in an oscillation of the kinetic energy around a value close to the initial energy, with \( \Delta {KE} = 0.001378 \). Given the near-zero value of \( \Delta {KE} \), this collision is nearly elastic, yet a noticeable width exchange between the solitons still occurs. Subfigure (c) presents an elastic collision with a velocity of \( v_0= -0.16 \), where \( \Delta {KE} \approx 0 \) and no width exchange takes place. Lastly, (d) showcases an inelastic collision with a velocity of \( v_0 = -0.28 \), accompanied by a width exchange, with a recorded \( \Delta {KE} = -0.281656 \). This leads to conclude that there are three possible outcomes for flat-top soliton collisions: 1) Completely elastic collisions with no radiation and no width exchange, 2) Nearly elastic collision with almost no radiation but with width exchange, and 3) Inelastic collision with both emitted radiation and width exchange. The rather surprising appearance of a single case of completely elastic scattering, imbedded within a continuum of inelastic collisions, is to be explained in the next section. It should be noted that, here and through out, the spacial domain is taken large enough such that the back-scattered radiation from the edges of the spacial grid do not reach the soliton during the evolution time.

\begin{figure}[htbp]
  \centering
  \includegraphics[width=8.5cm]{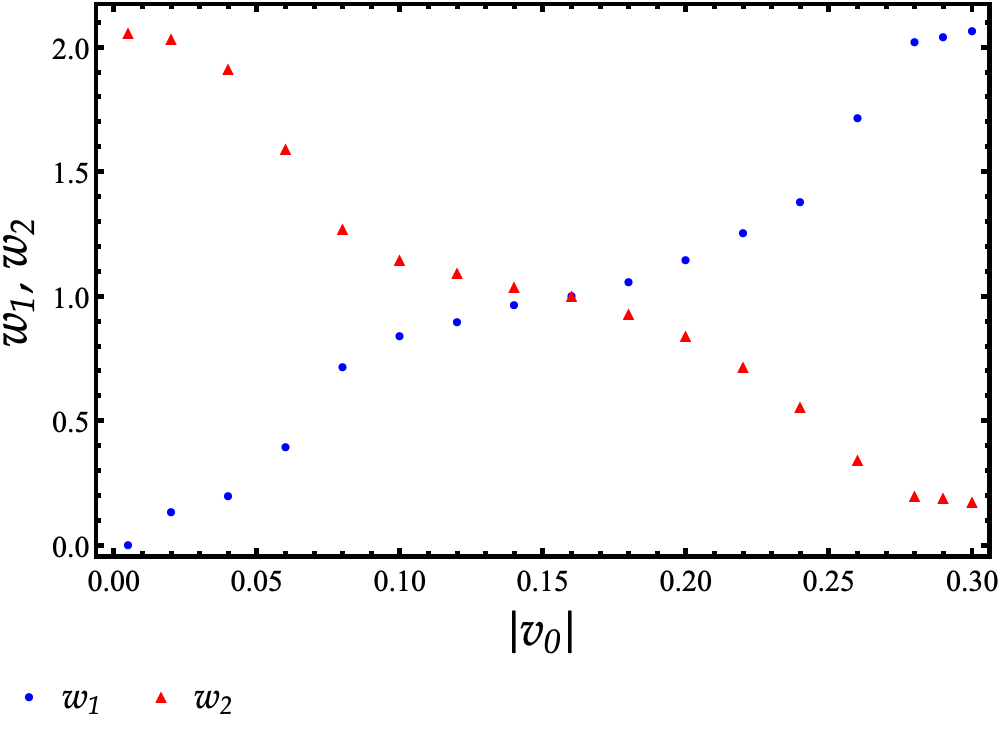}
\caption{Variation of flat-top solitons width ratios $w_1=W_1(t=t_f)/W_1(t=0)$ and $w_2=W_2(t=t_f)/W_2(t=0)$ versus $\gamma$. Here, $w_{1}$ and  $w_{2}$ denote the width ratio of the first and second solitons,  respectively. Values of other parameters are: $g_1=1/2,\,g_2=1,\,u_0=1$.}
  \label{fig5}
\end{figure}

To elucidate  quantitatively the phenomenon of width exchange between two flat-top solitons, as observed in Fig.~\ref{fig4}, we introduce the width ratios $w_{1}=W_1(t=t_f)/W_1(t=0)$ and $w_{2}=W_2(t=t_f)/W_2(t=0)$ to quantify the changes in their widths. Here, $w_{1}$ is defined as the ratio of the post-interaction width to the pre-interaction width for the left flat-top soliton, while $w_{2}$ follows the same formula for the right flat-top soliton. These width ratios are plotted versus soliton speed in Fig.~\ref{fig5}. From the figure, it is observed that at low velocities, the left flat-top soliton emerges from the collision with a reduced width compared to its original width, while the right flat-top soliton experiences the opposite effect. Upon increasing the velocity, a critical point is reached around \(|v_0|= 0.16\), at which no width exchange is observed. Past this point, the left flat-top soliton begins to emerge post-collision with an increased width. To comprehend this behavior, we list the values of \(\Delta {KE}\) as a function of velocity in Table~\ref{table1}. Notably, at \(|v_0|=0.16\), \(\Delta {KE}\) attains its minimum value, registering at a nearly negligible $-2.074\times10^{-7}$. Thus, indeed at this distinguished speed of  \(|v_0|=0.16\), the collision is completely elastic and no width exchange takes place.

\begin{table}[h!]
\centering
\caption{Change in kinetic energy, $\Delta {KE}$, radiation production,  $E_{rad}$, and width exchange expressed by the width ratios $w_1$ and $w_2$ of different flat-top soliton collisions. Values of other parameters are: $g_1=1/2,\,g_2=1,\,u_0=1$.}
\label{table1}
\begin{tabular}{ c|c|c|c } 
\hline
$|v_0|$& {$|\Delta \text{KE}|$}&$E_{rad}$ &$w_1,w_2$\\ 
\hline
0.005 & $1.494\times10^{-1}$& $2.956\times10^{-1}$ &$0.0000^*,2.0536$  \\
0.02  & $1.294\times10^{-1}$ & $9.981\times10^{-2}$&$0.1313,2.0322$  \\
0.04  & $1.243\times10^{-1}$ & $2.033\times10^{-4}$&$0.1964,1.9107$   \\
0.06  & $1.374\times10^{-2}$&$1.149\times10^{-7}$ &$0.3929,1.5893$  \\
0.08  & $2.567\times10^{-3}$ &$8.654\times10^{-9}$& $0.7143,1.2679$   \\
0.10  &$7.425\times10^{-7}$ &$4.428\times10^{-9}$ & $0.8393,1.1429$  \\
0.12  & $4.397\times10^{-7}$&$3.924\times10^{-9}$ &$0.8929,1.0893$  \\
0.14  & $4.107\times10^{-6}$&$1.009\times10^{-9}$& $0.9643,1.0357$  \\
0.16  &$2.074\times10^{-7}$& $5.726\times10^{-10}$ & $1.0000,1.0000$ \\
0.18  & $2.218\times10^{-5}$&$1.920\times10^{-8}$ & $1.0536,0.9286$  \\
0.20  & $1.443\times10^{-4}$& $3.686\times10^{-8}$ &$1.1429,0.8393$ \\
0.22  & $7.947\times10^{-4}$ &$7.642\times10^{-7}$& $1.2500,0.7143$  \\
0.24  &$8.219\times10^{-3}$ &$4.899\times10^{-5}$ &$1.3750,0.5536$ \\
0.26  & $5.893\times10^{-2}$&$1.402\times10^{-3}$ &$1.7143,0.3393$\\
0.28  & $2.729\times10^{-1}$ & $2.567\times10^{-2}$& $2.0398,0.1964$ \\
0.30   & $3.422\times10^{-1}$ & $1.874\times10^{-1}$& $2.0621,0.1712$ \\
\hline
\end{tabular}
\begin{tablenotes}
	\small
	\centering
	\item $^*$Two flat-top solitons combine after the collision to form a single flat-top soliton.
\end{tablenotes}
\end{table}

\begin{figure}[htbp]
  \centering
  \includegraphics[width=8.5cm]{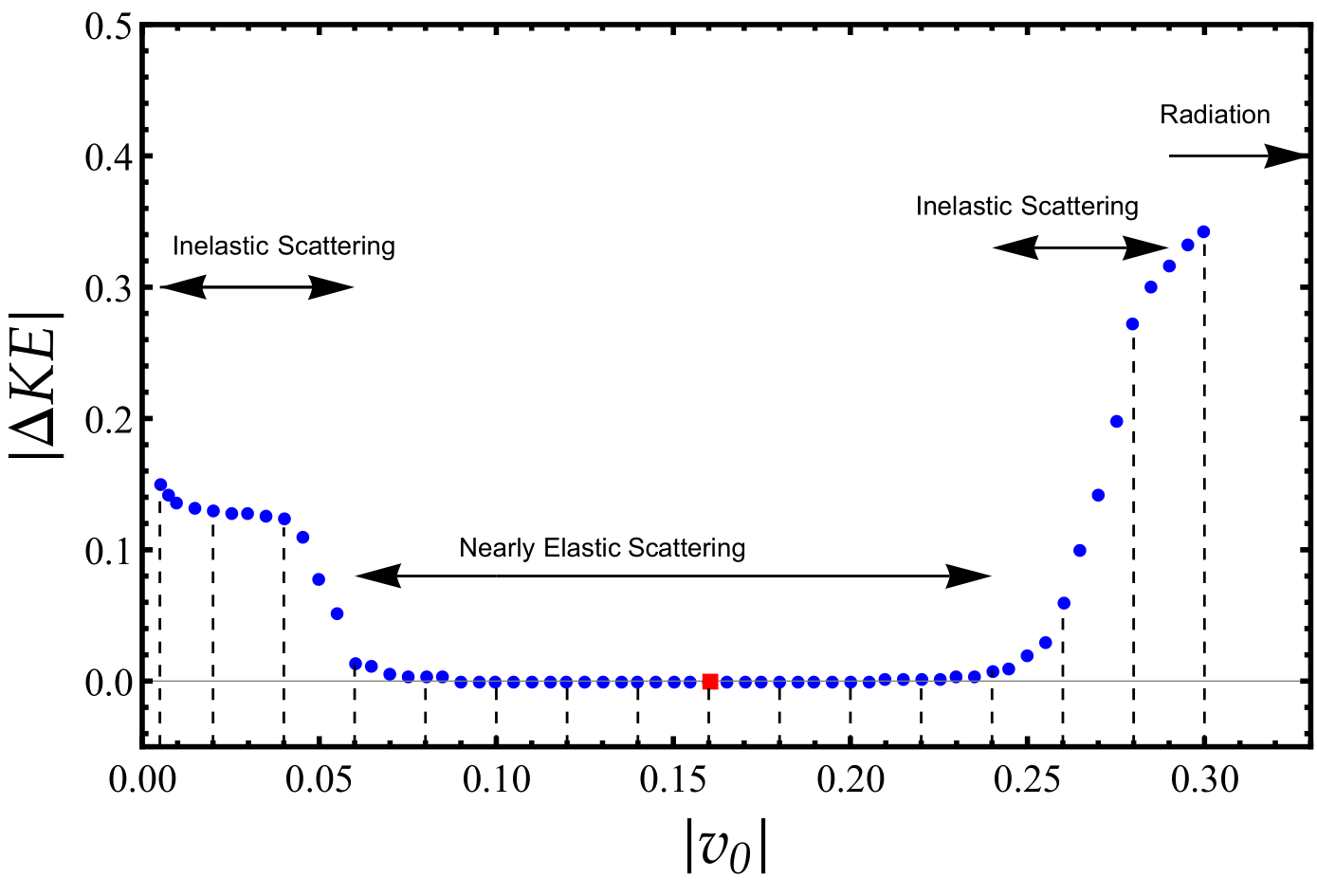}
  \caption{Inelastic, nearly elastic, and radiation regimes for the flat-top soliton collisions with width defined by $\gamma_{9}$.  A completely elastic collision, indicated by the red square, occurs at $|v_0|=0.16$. }
  \label{fig6}
\end{figure}

For an overview of the scattering outcomes, we identify in Fig.~\ref{fig6} the velocity ranges corresponding to elastic and inelastic domains, as well as regions exhibiting radiational effects, using the data presented in Table~\ref{table1}. For low velocities \( |v_0| < 0.08 \), the collision is inelastic. As the collision velocity increases, the interaction transitions to being nearly elastic for velocities in the range \( 0.08 \leq |v_0|\leq 0.22 \), marked by a reduction in the \(\Delta KE \) by two orders of magnitude. Nonetheless, width exchange is still ongoing in this regime, as shown in the table. For velocities above \( |v_0|= 0.22 \), we once again observe inelastic collisions. Notably, at relatively high collision velocities, \( |v_0| > 0.30 \), the inelastic collision not only results in a single flat-top soliton, but also gives rise to small-amplitude density waves indicative of radiation. It is noteworthy that at \(|v_0|\approx0.16\), the collision becomes fully elastic, marked by an absence of width exchange, absence of width oscillations, and absence of radiation. The numerical value for $\Delta KE$ is reduced, at this point, by anywhere from 2 to 5 orders of magnitude compared to the other data in Table~\ref{table1}. Strictly speaking, radiation is almost always produced in flat-top solitons scattering, but it is negligibly small in the region of nearly elastic scattering and is absent only in a single case of completely elastic scattering, as shown in Table~\ref{table1}. As will be discussed in Sec.~\ref{sub33}, the completely elastic scattering case is part of periodic occurrences at specific values of the soliton parameters. 

\begin{figure}[htbp]
	\centering
	\includegraphics[width=8cm]{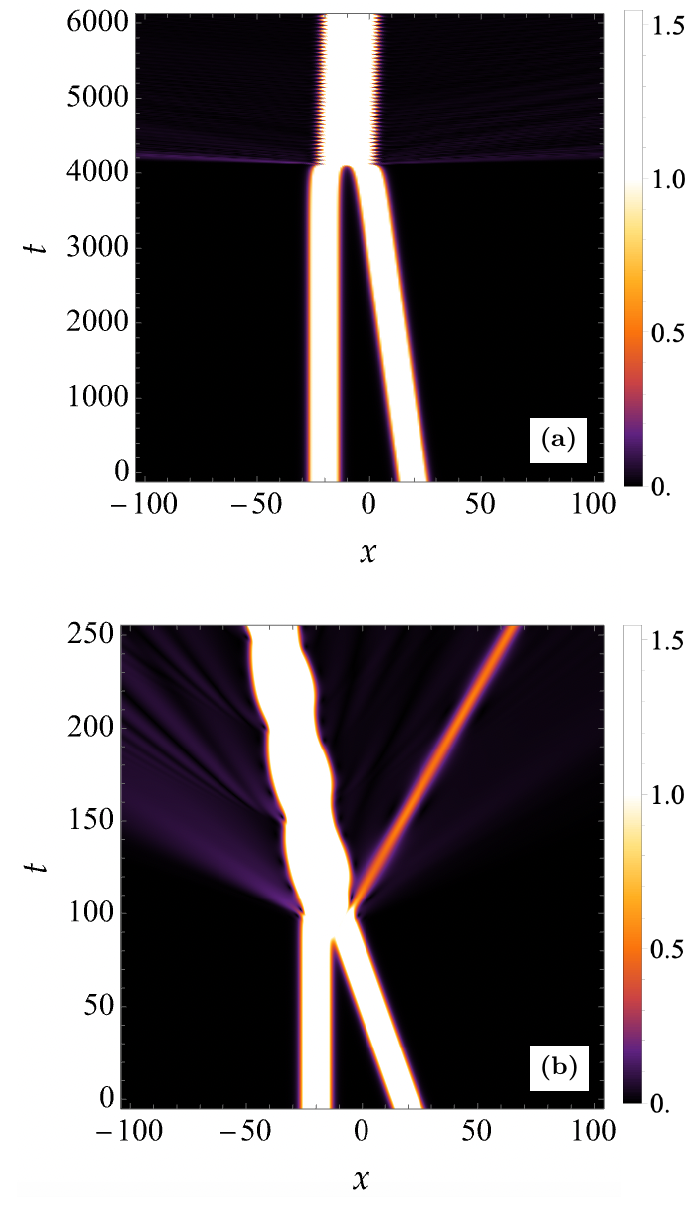}
	\caption{Two flat-top solitons at very low and very high velocities. A stationary flat-top soliton is approached by another flat-top soliton moving with (a) $v_0=-0.005$ and (b) $v_0= -0.30$. Spacial domain taken is $L=800$.}
	\label{fig7}
\end{figure}

The emergence of a single flat-top soliton at high velocity, along with the symmetry in the data in Table~\ref{table1}, i.e., \(\Delta {KE}\) is large at both low and high velocities due to the ejection of small density waves, raises the question of whether two flat-top solitons might similarly merge at very low velocities. Figure~\ref{fig7} depicts this phenomenon, showing two flat-top solitons combining into a single entity at both very low \((|v_0|=0.005)\) and high \((|v_0|=0.30)\) collision velocities. 

\begin{figure}[htbp] 
	\centering	
	\includegraphics[width=8.5cm]{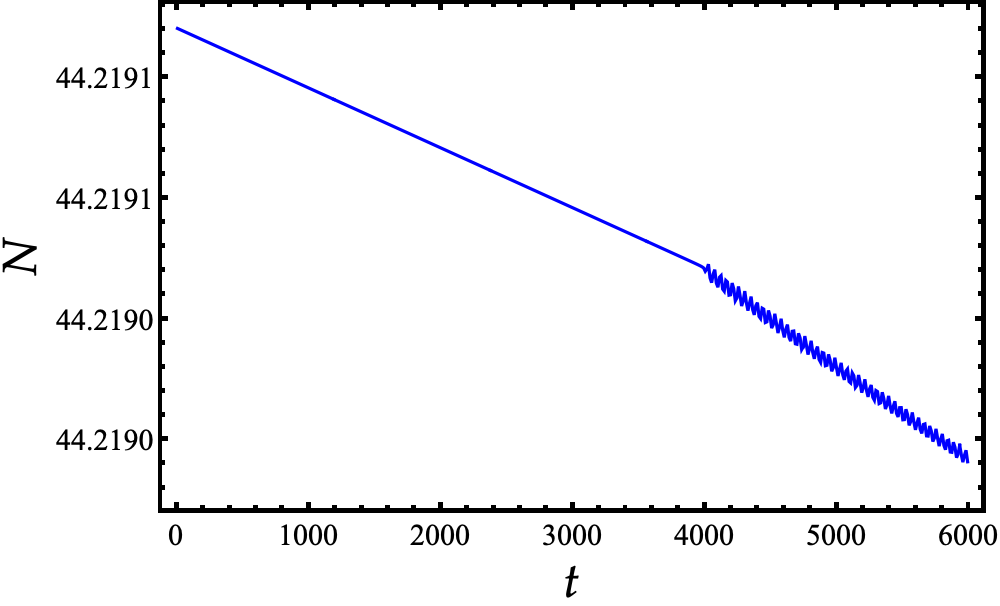}\\
	\includegraphics[width=8.5cm]{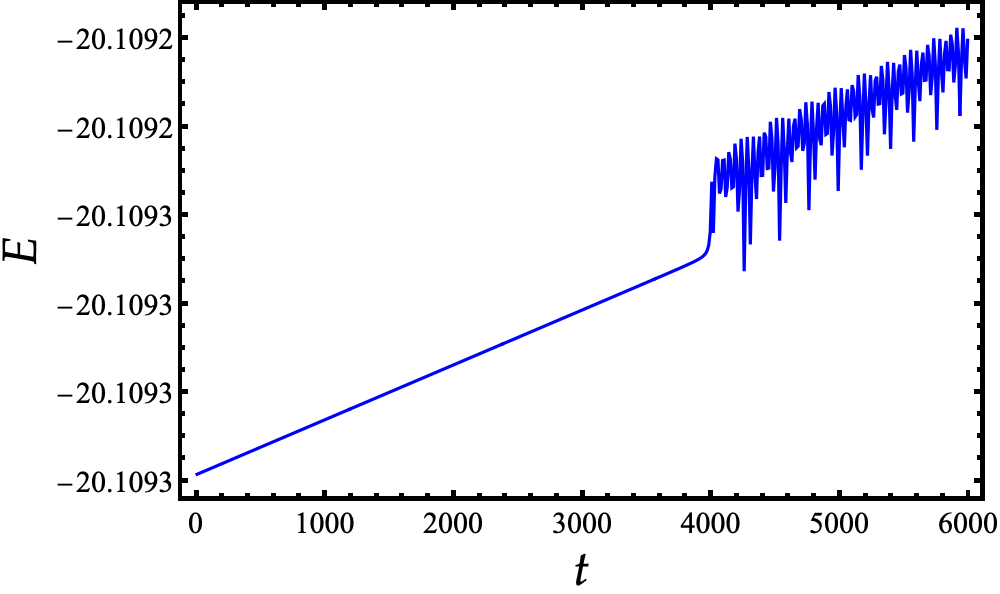}\\
	\includegraphics[width=8.5cm]{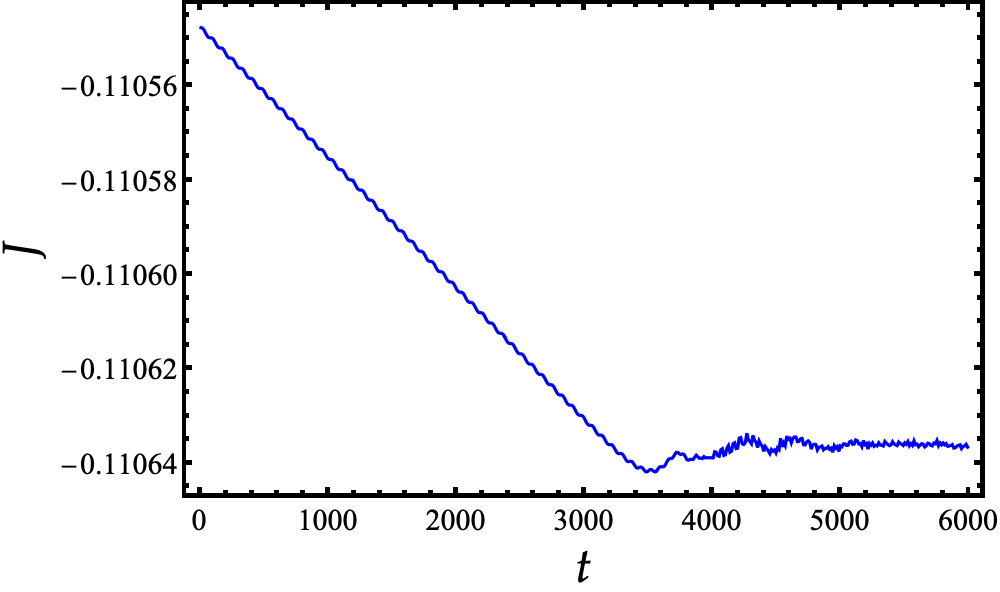}
	  \begin{picture}(5,5)(5,5)
	\put(-25,435) {\color{black}{{\fcolorbox{white}{white}{\textbf{(a)}}}}}
	\put(-58,277) {\color{black}{{\fcolorbox{white}{white}{\textbf{(b)}}}}}
	\put(-25,132) {\color{black}{{\fcolorbox{white}{white}{\textbf{(c)}}}}}
	\end{picture}	
	\caption{Time evolution of norm, (a), total energy, (b), and momentum, (c) of the flat-top soliton shown in Fig. \ref{fig7}(a), showing their conservation, within the accuracy of the numerical method.}
	\label{fig8}
\end{figure}
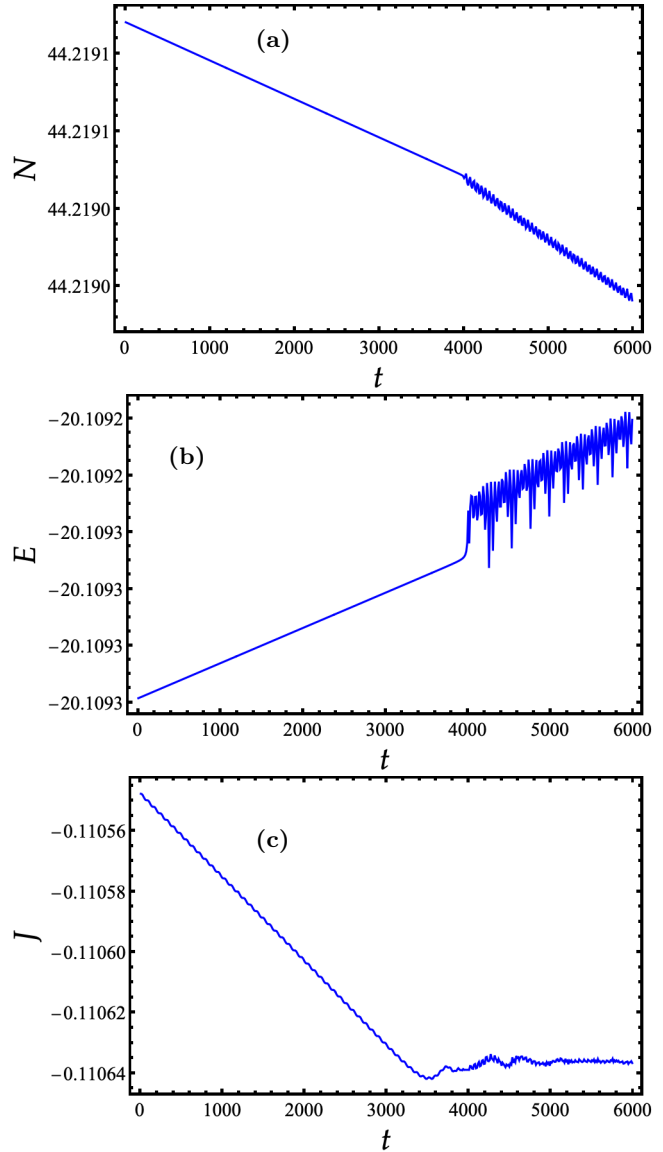

\begin{figure}[!ht]    
\includegraphics[width=8.5cm]{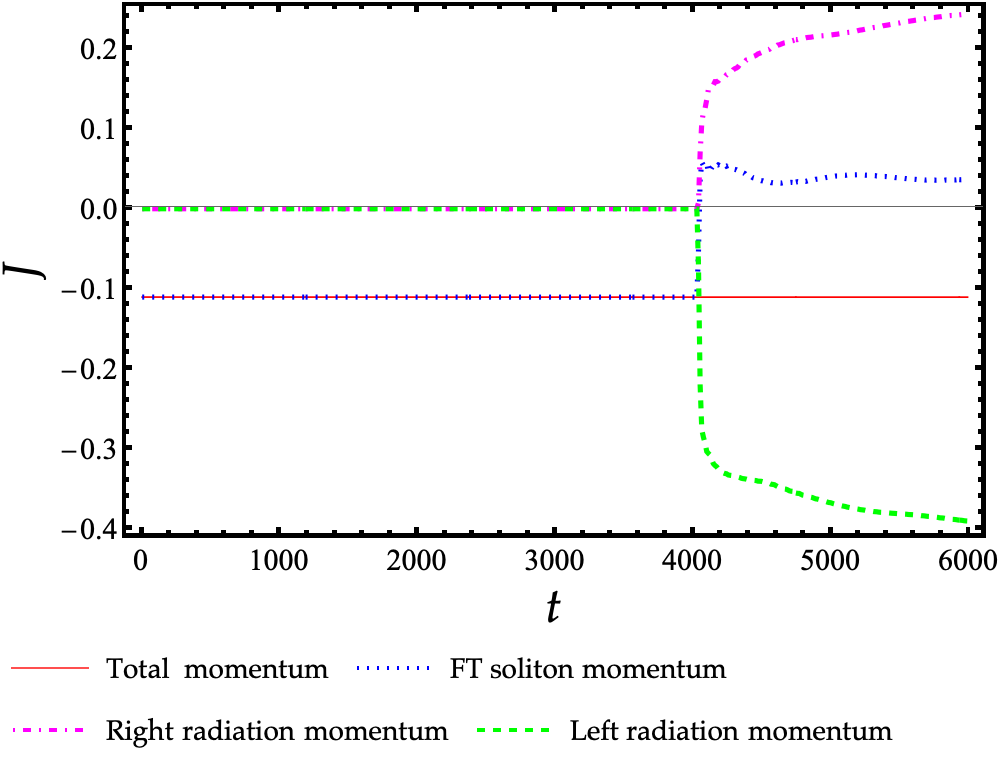}
\caption{Time evolution of the momentum of the coalesced flat-top soliton and the associated radiation for  the scattering shown in Fig. \ref{fig7}(a).} 
\label{fig9}
\end{figure}

It  is interesting to observe that the coalesced soliton acquires a positive momentum in spite of that the total initial momentum is negative, which raises a question about the conservation of total momentum. To understand this unusual behavior, we first verify that total energy, norm, and momentum are indeed conserved, up to the accuracy of the numerical method, as shown in Fig.~\ref{fig8}. Since total momentum
\begin{equation}
J=\frac{i}{2}\,\int_{-\infty}^{\infty}\left(\frac{\partial\psi^*(x,t)}{\partial x}\,\psi(x,t)-\frac{\partial\psi(x,t)}{\partial x}\,\psi^*(x,t)\right)dx
\label{mom}
\end{equation}
is conserved, the total momentum before collision is negative, the momentum of the coalesced solitons after collision is positive, then some negative momentum should be carried by the scattered radiation for the total momentum to be conserved. We have calculated the momentum carried by  the radiation part and the momentum carried by the solitons separately, as shown in Fig.~\ref{fig9}. The figure shows that the momentum of the radiation scattered to the left is larger  than the momentum of radiation scattered to  the right, such that  the total momentum carried by radiation is indeed negative and total momentum  of both coalesced soliton and radiation is consequently conserved. 

\subsection{Collision Behaviors of Flat-Top Solitons Across Widths Spaing \(\gamma_{0}\) to \(\gamma_{16}\)}
\label{sec:FTS_gamma0To16}

\begin{figure}[htbp]
  \centering
  \includegraphics[width=8.5cm]{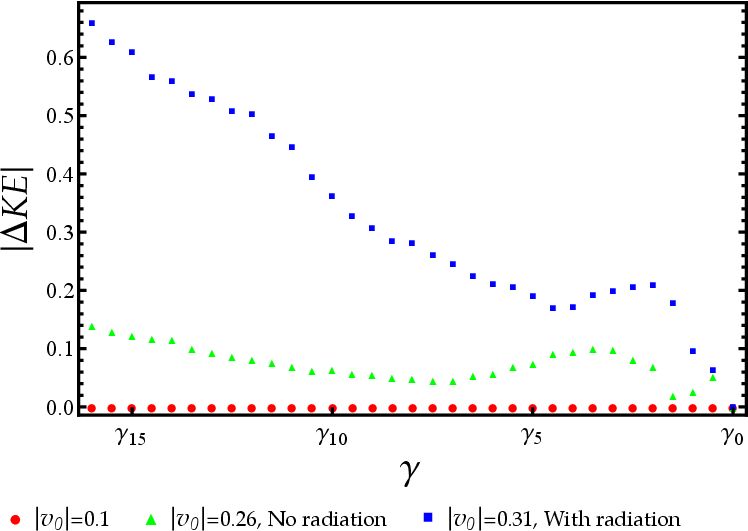}
  \caption{Variation of the change in kinetic energy, $|\Delta KE|$, and $\gamma$ for three flat-top soliton collisions, each characterized by the velocity of the moving flat-top soliton. Values of other parameters are: $g_1=1/2,\,g_2=1,\,u_0=1$.}
  \label{fig10}
\end{figure}

It is instructive to investigate the elastic collisions between flat-top solitons of varying sizes. To achieve this, we conducted numerical calculations to outline the regions corresponding to elastic, inelastic, and radiative behaviors for flat-top solitons with widths corresponding to the range \( \gamma_{0}-\gamma_{16} \). We selected three representative velocity values to characterize the elastic, inelastic, and radiation regions: \( |v_0|=0.1 \), \( |v_0|=0.26 \), and \( |v_0|=0.31 \), as illustrated in Fig.~\ref{fig6}. In Fig.~\ref{fig10}, we utilize these velocity values to plot $\Delta KE$ versus $\gamma$. For bright solitons (\( \gamma_{0} \)), our findings confirm that the kinetic energy remains constant before and after the collision, implying purely elastic behavior for all velocities. It is worth noting that bright solitons are well-known for their exclusively elastic collision properties, as discussed above. In contrast, for \( \gamma\ne0 \) we identify three distinct regions based on collision behavior. At \(|v_0|=0.1\), although there is width exchange, all collisions are nearly elastic, as \(\Delta KE \approx 0\); at \(|v_0|=0.26\), they are inelastic, given that \(\Delta KE \neq 0\). The degree of inelasticity varies with the width of the flat-top solitons. At \( |v_0|=0.31 \), we observe an inelastic collision that is also accompanied by radiation. The findings indicate a notable divergence in the collisional properties of flat-top solitons compared to the traditional behaviors observed in the scattering dynamics of two bright solitons.

\begin{figure}[htbp] 
  \centering
    \includegraphics[width=8.5cm]{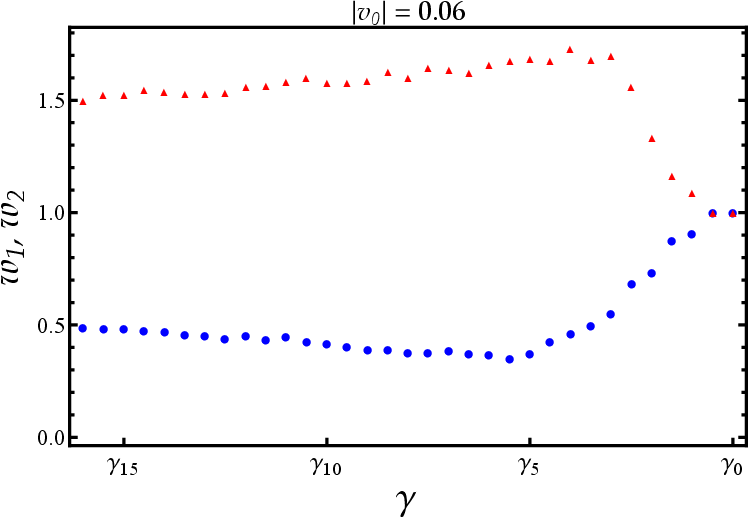}\\
    \includegraphics[width=8.5cm]{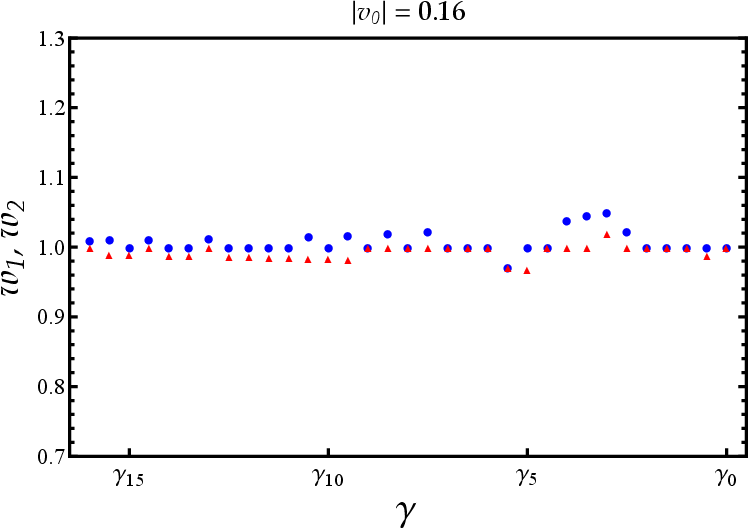}\\
    \includegraphics[width=8.5cm]{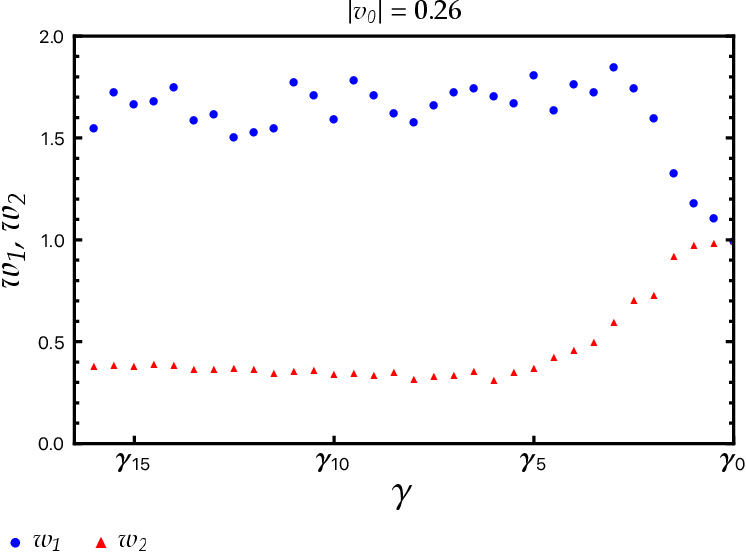}
    \begin{picture}(5,5)(5,5)
    \put(-75,515) {\color{black}{{\fcolorbox{white}{white}{\textbf{(a)}}}}}
    \put(-75,335) {\color{black}{{\fcolorbox{white}{white}{\textbf{(b)}}}}}
    \put(-78,172) {\color{black}{{\fcolorbox{white}{white}{\textbf{(c)}}}}}
    \end{picture}  
  \caption{Variation of width ratios of flat-top solitons with $\gamma$ from $\gamma_{0}$ to $\gamma_{16}$ for three flat-top soliton collisions, each characterized by the velocity of the moving flat-top soliton. Here, $w_{1}$ and  $w_{2}$ denote the width ratio of the first and second solitons,  respectively. (a) $|v_0|=0.06$, (b) $|v_0|=0.16$, and  (c) $|v_0|=0.26$. Values of other parameters are: $g_1=1/2,\,g_2=1,\,u_0=1$.}
  \label{fig11}
\end{figure}
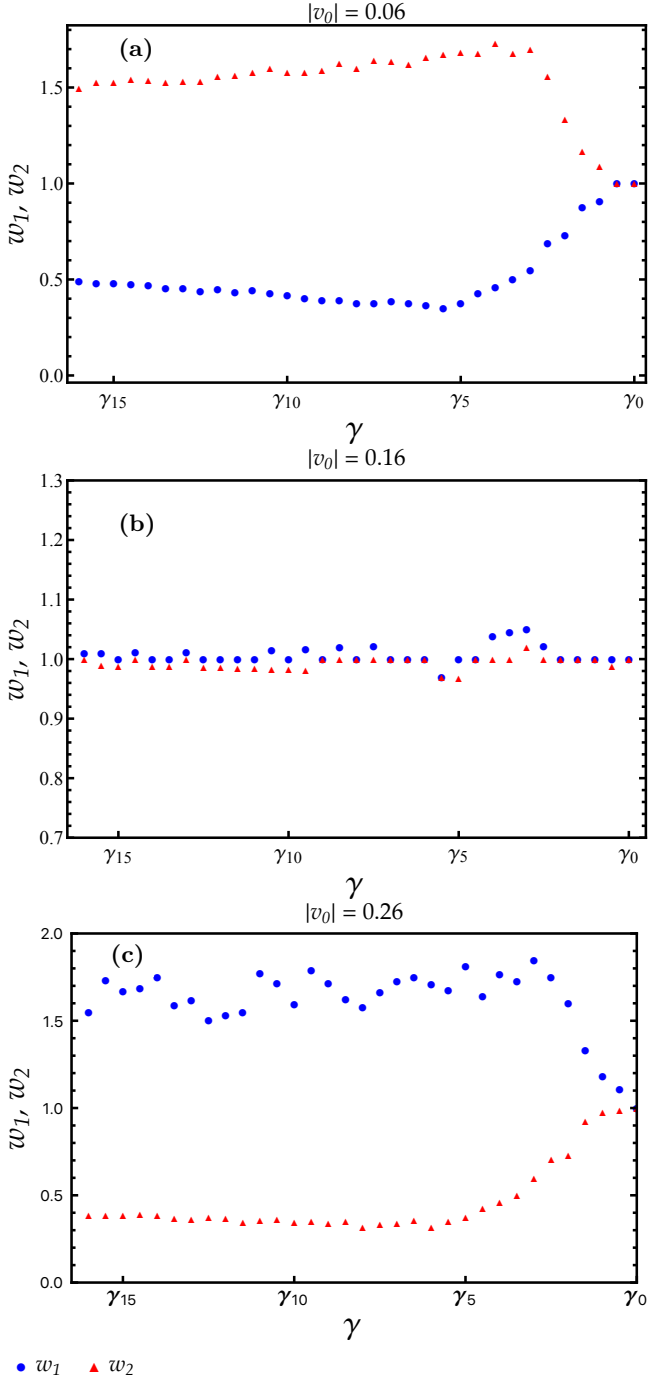

In Fig. \ref{fig11}, we examine the exchange of widths that occurs when two flat-top solitons collide with widths corresponding to the range $\gamma_0-\gamma_{16}$. Referring to Fig.~\ref{fig4}, a significant width exchange is evident at both low and high velocities. We investigated this phenomenon at three specific velocities: \(|v_0|=0.06\), \(|v_0|=0.16\), and \(|v_0|=0.26\). According to Fig.~\ref{fig4}, there appears to be no width exchange around \(|v_0|=0.16\), prompting us to investigate whether this holds true for flat-top solitons of various widths.
In Fig.~\ref{fig11}(a), we observe that increasing the flat-top soliton width leads to an exchange in widths following the collision. Notably, the rate of width exchange stabilizes for flat-top solitons with widths corresponding to $|\gamma|>|\gamma_6|$. At this velocity, there is an absence of a noticeable width exchange for flat-top solitons with a width corresponding to $\gamma_1$.
Figure~\ref{fig11}(b) presents a fascinating finding: there is minimal width exchange for all flat-top solitons of different widths, as \( w_{1,2}\) is nearly equal to 1. This outcome confirms that at this velocity (\(v_0=-0.16\)), the collision behavior of flat-top solitons resembles that of bright solitons, characterized by the absence of width exchange and a situation where \(\Delta KE = 0\).
For \(|v_0|=0.26\), as shown in Fig.~\ref{fig11}(c), we observe results akin to those in Fig.~\ref{fig11}(a), with a width exchange also taking place. However, there is a distinct difference: while the left flat-top soliton had a smaller width after the collision at \(|v_0|=0.08\), it now exhibits a larger width after the collision.


\section{Periodic Occurrence of Elastic Scattering}
\label{sub32}
\begin{figure*}[htbp]
  \centering
    \includegraphics[width=\textwidth]{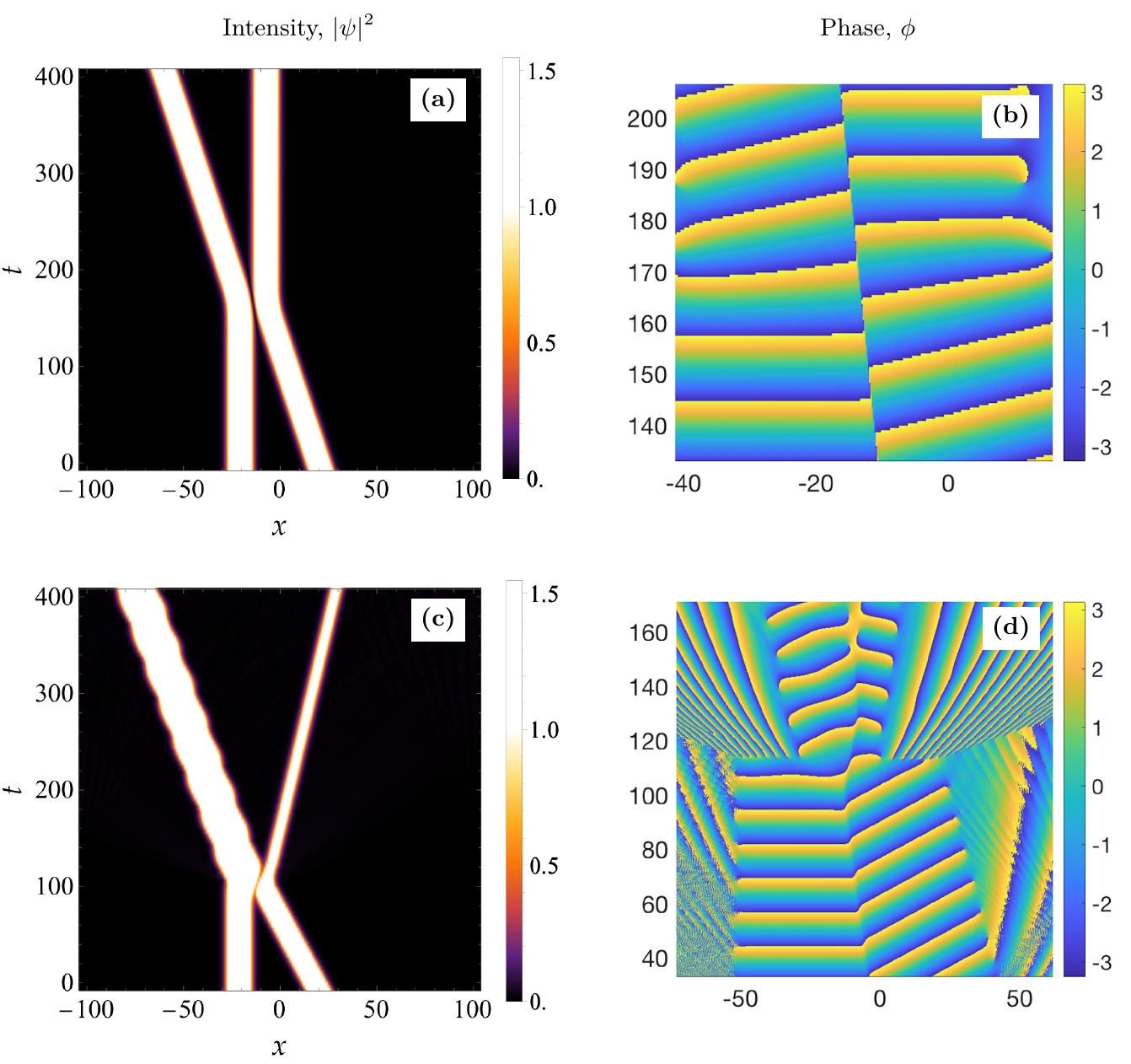}
  \caption{Density and phase plots of collisions between a stationary flat-top soliton and a moving flat-top soliton, respectively. (a), (b) elastic collision with $v=0.16$ and (c), (d) inelastic collision $v=0.26$. Values of other parameters are: $g_1=1/2,\,g_2=1,\,u_0=1$.}
  \label{fig12}
\end{figure*}

We found in the previous section that completely elastic collision with no width exchange occurs only at a specific value of the initial soliton speed. Attempting to find an explanation for this behavior, we observe in Fig.~\ref{fig4}(c) that flat-top solitons do not overlap at this case of completely elastic scattering. Invoking the well-established fact that solitons of $\pi$-phase difference repel each other \cite{gordon}, we conjecture that completely elastic collision without width exchange occurs when the two flat-top solitons have a $\pi$-phase difference just before their collision. In Fig.~\ref{fig12}, this is confirmed by plotting the phase for completely elastic case and one inelastic case. The present section is devoted for further proofs and consequences of this conjecture.

The first test of our conjecture is to repeat the completely elastic scattering case of Fig.~\ref{fig4}(c) (or equivalently Fig.~\ref{fig12}(a)), namely with $|v_0|=0.16$, but with a constant phase, $\phi_0$, added to the  moving soliton. The initial profile, in this case, takes the form
\begin{equation}
\psi(x,0)=\psi_l(x,0)+\psi_r(x,0)\,e^{i\phi_0},
\end{equation}
where $\psi_l(x,0)$ is the initial profile of the left (stationary) soliton, as given by Eq.~(\ref{ftsol}) with $t=0$ and $v_0=0$, and  $\psi_r(x,0)$ is the initial profile of the right (moving) soliton, as given by Eq.~(\ref{ftsol}) with $t=0$. 

\begin{figure}[!ht]    
\includegraphics[width=8.5cm]{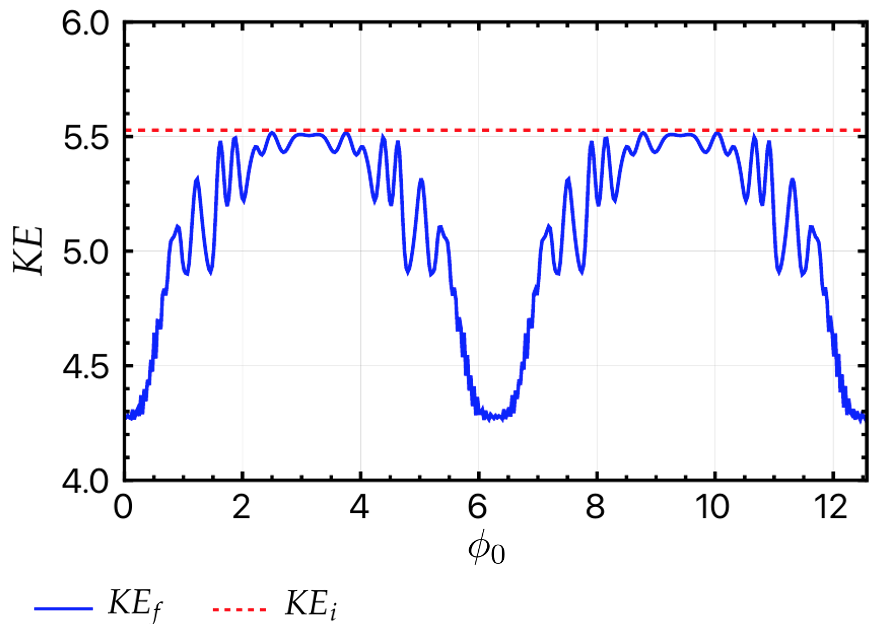}
\caption{Kinetic energy after collision of two flat-top solitons (solid blue) and their kinetic energy before collision (dashed red), in terms of the initial relative phase, $\phi_0$. Parameters used: $g_1=1/2,\,g_2=1,\,u_0=1, v_0=-0.64,\,x_0=20.0$.} 
\label{fig13}
\end{figure}

If our conjecture is correct, we should see periodic appearance of completely elastic scattering in terms of $\phi_0$ at values of relative phase $\phi_0=\pi+2m\pi$, where $m$ is an integer. Indeed, this is what we see in Fig.~\ref{fig13}, where a completely elastic collision is attained for $\phi_0=\pi$ and $\phi_0=3\pi$. For values of $\phi_0\ne\pi+2m\pi$ an overlap between the two flat-top solitons occurs, and whenever there is an overlap, there is a kinetic energy loss. It is observed  that the maximum energy loss occurs at $\phi_0=0,\,2\pi,\,4\pi$, which correspond to in-phase solitons. It is only when the two solitons are completely out-of-phase, $\phi_0=\pi+2m\pi$, the repulsion between the two flat-top solitons is strong enough such that they do not  overlap and thus do not loose any of their initial kinetic energy. This connection between overlap and loss in kinetic energy will be helpful to understand the physics underlying the inelasticity, discussed in more details in the next section.  

\begin{figure}[!h]    
	\includegraphics[width=8.3cm]{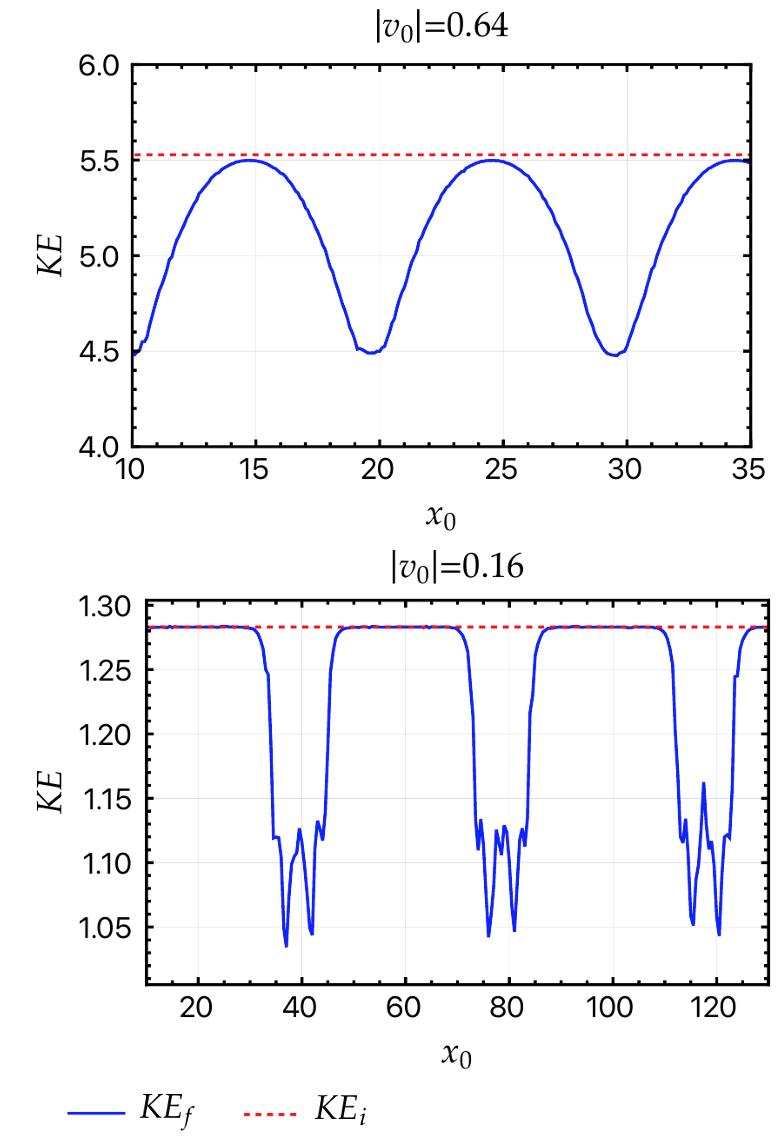}
	\caption{Kinetic energy after collision of two flat-top solitons (solid blue) and their kinetic energy before collision (dashed red), in terms of the initial separation between flat-top solitons, for two values of the initial relative speed. Parameters used: $g_1=1/2,\,g_2=1,\,u_0=1,\,\phi_0=0$. } 
	\label{fig14}
\end{figure}

Another evidence manifesting the dependence of collision elasticity on the relative phase is its dependence on the initial separation between the two flat-top solitons. Since phase is a dynamic quantity that evolves with time, it is expected that the phase at which the two flat-top solitons meet to be function of the distance that the initial soliton had to cross from the initial time till the time of the collision. We fix the left soliton and change the initial position of the right soliton and then calculate the kinetic energy change in terms of the initial separation. Plotting in Fig.~\ref{fig14} the change in kinetic energy in terms of the initial separation, we see the periodic appearance of the completely elastic scattering, as anticipated. The positions at which completely elastic scattering occurs, are $x_0=14.7,\,24.6,\,34.4$. The difference between any two consecutive values of these positions, say the first two, is about $\Delta x_0\approx10.0$. Our  conjecture that elastic collision occurs when the relative phase between the two solitons equals $\pi$ plus a multiple of $2\pi$ can now be verified by explaining two features of this curve. First, the distance between initial positions of any two successive elastic collisions should correspond to a phase change that is a multiple of $2\pi$. Secondly, the initial separation between the two solitons for any of the elastic scattering cases should correspond  to a $\pi$-phase difference when the two solitons collide, plus a multiple of $2\pi$. The phase of the incident soliton (on the right side), given by Eq. (\ref{ftsol}), is
\begin{equation}
\phi_r(x,t)= u_{0}t + {v_{0}\left(2(x - x_{0}) - v_{0}t\right)}/(4 g_{1}),
\end{equation} 
and the phase of the stationary soliton (on the left side) is
\begin{equation}
\phi_l(x,t)= u_{0}t.
\end{equation} 
It should be mentioned that these expressions describe almost exactly the evolution of phase for both solitons since the overlap between  them is negligible, except just before the  collision time. Therefore, these expressions are expected to give an accurate estimate of the relative phase between the two solitons just before they collide. To explain the first feature, namely the periodic appearance of elastic collision, the change in the relative phase, $\Delta(\delta\phi)$, due to a change in the position of the right soliton, $\Delta x_0$, must be a multiple of $2\pi$. Here, $\delta\phi$ is defined as the relative phase
\begin{equation}
\delta\phi=\phi_r(x,t)-\phi_l(x,t)=\frac{v_0(2(x-\,x_0)-v_0t)}{4g_1}.
\end{equation} 
The two solitons meet at $x=-x_0$ at a time $t=-2x_0/v_0$. Their relative phase when they meet is thus
\begin{equation}
\delta\phi(x_0,v_0)=\frac{x_0v_0}{2g_1}\label{dph}.
\end{equation} 
When the initial position is changed from $x_{01}$ to $x_{02}$, where $x_{01}$ and $x_{02}$ correspond to two elastic scattering cases, the phase difference changes as 
\begin{eqnarray}
\Delta(\delta\phi)&=&\delta\phi(x_{02},v_0)-\delta\phi(x_{01},v_0)\nonumber\\&=&\frac{\Delta x_0\,v_0}{2g_1}
\label{dph10},
\end{eqnarray} 
where $\Delta x_0=x_{02}-x_{01}$. 
For the first period of the curve in Fig.~\ref{fig14} for $|v_0|=0.64$, we read off $\Delta x_0=24.6 - 14.7$, which corresponds to a phase difference, according to (\ref{dph10}), that equals $\Delta(\delta\phi)=1.008411\times2\pi$. For the second period with $\Delta x_0=34.4 - 24.6$, the phase difference is  $\Delta(\delta\phi)=0.99822\times2\pi$. The second feature to be confirmed is that values of $x_0$ at minimum kinetic energy loss (peaks of the curves) should correspond to relative phase, $\delta\phi$, that equals $\pi$. For the present case, Eq.~(\ref{dph}) gives $\delta\phi=2.99466\times\pi,\,5.01147\times\pi,\,7.00791\times\pi$, corresponding peak positions at $x_0=14.7,\,24.6,\,34.4$, respectively. 

For further confirmation, we repeat this calculation for a soliton initial speed of $|v_0|=0.16$. In Fig.~\ref{fig14} for $|v_0|=0.16$,  the loss in kinetic energy for this case is closest to the initial kinetic energy at $x_0\approx20,\,59.5,\,100$, which correspond to relative phases $\delta\phi=1.01859\times\pi,\,3.03031\times\pi,\,5.093\times\pi$, respectively. Clearly, the out-of-phase condition for elastic scattering is also verified here as well.

\begin{figure}[!h]    
\includegraphics[width=8.5cm]{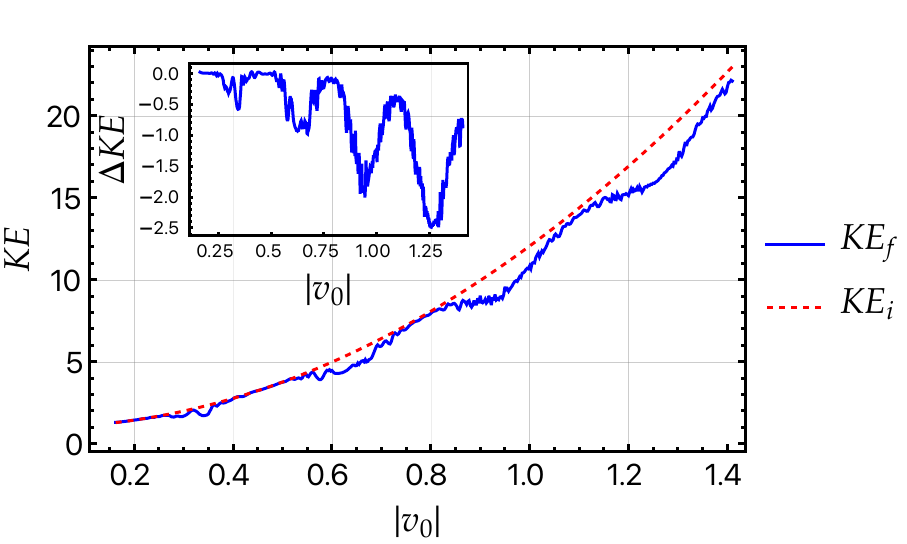}
\caption{ Kinetic energy after collision of two flat-top solitons (solid blue) and their kinetic energy before collision (dashed red), in terms of their initial relative speed, which equals the speed of the right soliton. Inset shows the kinetic energy change. Parameters used: $g_1=1/2,\,g_2=1,\,u_0=1, x_0=20.0,\,\phi_0=0$.  } 
\label{fig15}
\end{figure}

Initial soliton speed is another parameter that affects the relative phase of solitons, as seen by Eq.~(\ref{dph}). Therefore, as an additional proof on the role of the phase, we fix the initial soliton position and study the effect of initial soliton speed on elasticity. In Fig.~\ref{fig15}, we plot kinetic energy versus initial soliton speed, $v_0$. Indeed, periodic appearance of elastic scattering occurs at speeds $|v_0|=0.470,\,0.785,\,1.100$. Equation  (\ref{dph10}) gives the phase difference in terms of the change in the initial speed as follows: 
\begin{equation}
\Delta(\delta\phi)=\frac{x_0\,\Delta v_0}{2g_1}.
\end{equation}  
Relative phase corresponding to the above initial soliton speeds are: $2.99211\times\pi,\,4.99747\times\pi,\,7.00282\times\pi$, which correspond  to out-of-phase solitons and hence provide another evidence that elastic scattering occurs when  solitons are out-of-phase. 

\section{Energy Transfer Analysis: Origin of Inelasticity}
\label{sub33}
\begin{figure}[!h]    
\includegraphics[width=8.5cm]{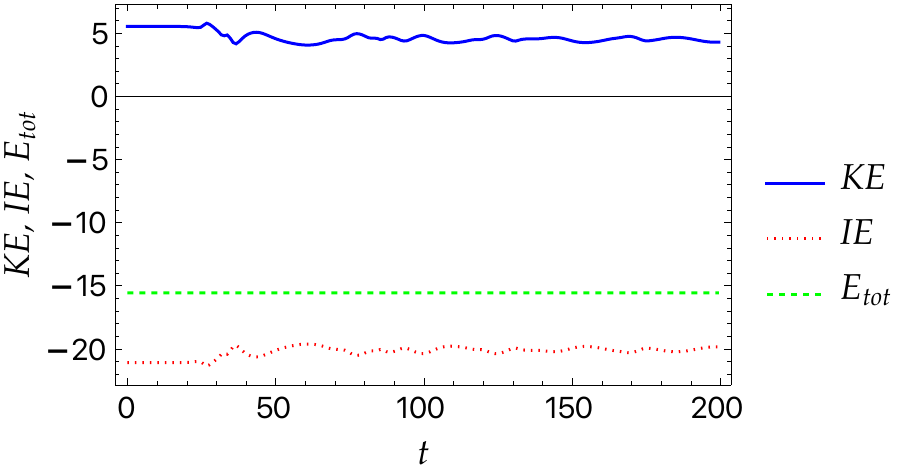}
\caption{ Kinetic energy, $KE$, interaction energy, $IE$, and total energy, $E_{tot}=KE+IE$, including both solitons and radiation. Parameters used: $g_1=1/2,\,g_2=1,\,u_0=1, v_0=-0.64$.  } 
\label{fig16}
\end{figure}

Loss in the kinetic energy signifies inelasticity in soliton scattering, as shown in Fig.~\ref{fig16}, which corresponds to a completely inelastic scattering where the two solitons remain to be coalesced  after scattering. The  reduction in the total kinetic energy is balanced by an equal increase in the interaction energy, such that  the total energy  is constant.  However, this energy transfer does not explain why the two solitons continue to be coalesced and do not retain their state of two separate solitons as bright solitons do after coalescence. Scattering of bright solitons is completely elastic although solitons do coalesce momentarily and part of the kinetic energy transfers to interaction energy, bright solitons have somehow an `internal memory' such that they retain their status before collision of two separate solitons with exactly the same kinetic energy. The question we try to answer here is: why this does not happen to flat-top solitons? why flat-top solitons do not retain their state of two separate solitons as before scattering?  As, we have seen in the previous section, flat-top soliton scatterings are always inelastic, except for cases when the two solitons are completely out of phase at the moment of scattering, where the force of repulsion is strong enough to prevent the solitons from overlapping. Inelasticity occurs always only when the profiles of the two solitons overlap. So, we rephrase the above questions as: when two flat-top solitons scatter, why  the reduction in the kinetic energy, which is supposed to transfer momentarily to interaction energy, is not retained? Had solitons retained their original kinetic energy, they will have the momentum to separate after coalescence, but with a kind of `permanent' loss in the kinetic energy, the interaction energy of the coalesced solitons overcomes the kinetic energy such that the two solitons are `locked' into this state of coalescence. 
The followup question is what causes this permanent loss in the kinetic energy? why in bright solitons scattering it seems  there is no such permanent loss in the kinetic energy? Our inspection showed that the key answer to these questions is in radiation production. Emitted radiation carries with it some kinetic energy and momentum, which are permanently lost from the solitons since radiation is a dispersive type of wave and propagates faster than solitons. Bright solitons, on the other hand, do not emit radiation during their scattering since they are exact solutions to a completely integrable system, namely the fundamental NLSE, i.e. Eqs.~(\ref{cqnlse}) and (\ref{ftsol}) with $g_3=\gamma=0$. Integrable systems are characterized by an infinite number of conserved quantities. The NLSE with dual power law nonlinearities is not integrable and hence radiation is always produced during the scattering of its exact solutions, which leads to an irreversible loss in kinetic energy and hence inelastic scattering of flat-top solitons.  

To verify this explanation, we perform a detailed analysis to the energy transfers of the different energy components. We take the specific case of  $\gamma=\gamma_9=10^{-9}-1$ corresponding to left-moving soliton with velocity  $v_{0}=-0.64$ scattered by a stationary soliton, leading to a coalesced single soliton accompanied by radiation. The energy components of solitons and radiation are calculated before and after scattering separately, as listed in Table~\ref{table2}. Before scattering, energies are calculated at the initial time. After scattering, the energies are recorded after an evolution time sufficient to reach a steady state.  Center-of-mass kinetic energy, $KE_{com}=4.52794$, is carried initially by the left-moving soliton; the other soliton is initially stationary. Total kinetic energy pressure, $KE_{pressure}=0.999485$, is the sum of two equal kinetic energy pressure values of the two solitons. Total kinetic energy is the sum of these two kinetic energy components: $KE_{tot}=KE_{com}+KE_{pressure}=5.52743$. After collision, $KE_{com}$ reduces to $2.10299$,  $KE_{pressure}$ reduces to $0.464666$, and thus $KE_{tot}$ reduces to $2.56765$. Interaction energy, on the other hand, increases from $-21.1096$ to $-19.6569$. Total kinetic energy decreases by $2.95977$ while interaction energy increases only by $1.45270$. Thus, out of the total reduction in kinetic energy, $2.95977$, only $1.45270$ transfers to interaction energy while  the rest, $2.95977-1.45270=1.50758$, transfers to radiation. Examining  the Radiation column in the table, we see that a total kinetic energy of $1.7113$ is gained by radiation accompanied by $-0.20418$ interaction energy. The  total energy carried by radiation is thus $1.71133 - 0.20418=1.50785$, which equals to the energy reduction by solitons kinetic energy. Another important feature to point out is that most of the radiation energy is center-of-mass kinetic  energy. This shows that about 50\% of  the reduction in the kinetic energy of the solitons transfers mainly to a center-of-mass kinetic energy of the radiation. Norm of solitons reduces from $44.2191$ for the two incident solitons to $41.2722$ for the coalesced solitons. The reduction in norm equals  the radiation norm, $2.9469$.  Center-of-mass speed of the incident  soliton is $0.64$, center-of-mass speed of the coalesced solitons is $0.319231$, and the effective center-of-mass speed for radiation is $1.0371$, which is almost 3 times faster than the speed of the coalesced solitons. This analysis, shows that reduction in the solitons kinetic energy is transferred to center-of-mass kinetic energy of radiation, which is never retained by the solitons due to the dispersive nature of radiation. 
\begin{table*}[bt]
	\centering
	\caption{Energy components  of solitons and radiation before and after scattering.}
	\label{table2}
	\centering
	\begin{tabular}{ l|lll|llll } 
		\hline
		\text{Energy components} & \text{} & \text{Soliton} & \text{} & \text{} & \text{} &
		\text{Radiation} & \text{} \\\hline
		\text{} & \text{Initial} & \text{} & \text{Final} & \text{} &
		\text{Initial} & \text{} & \text{Final} \\\hline
		$KE_{com}$ & 4.52794 & \text{} & 2.10299 & \text{} & 0 & \text{} &
		1.58479 \\\hline
		$KE_{pressure}$ & 0.999485 & \text{} & 0.464666 & \text{} & 0 &
		\text{} & 0.126536 \\\hline
		$KE_{tot}$& 5.52793 & \text{} & 2.56765 & \text{} & 0 & \text{} &
		1.71133 \\\hline
		$IE$ & -21.1096 & \text{} & -19.6569 & \text{} & 0 & \text{} &
		-0.20418 \\\hline
		$N$& 44.2191 & \text{} & 41.2722 & \text{} & 0 & \text{} &
		2.9469 \\\hline
		$v_{com}=\sqrt{{2KE_{com}}/{N}}$& 0.639998 & \text{} & 0.319231 & \text{} & 0 & \text{} &
		1.0371 \\\hline
	\end{tabular}
\end{table*}

\begin{figure}[!h]    
\includegraphics[width=8.5cm]{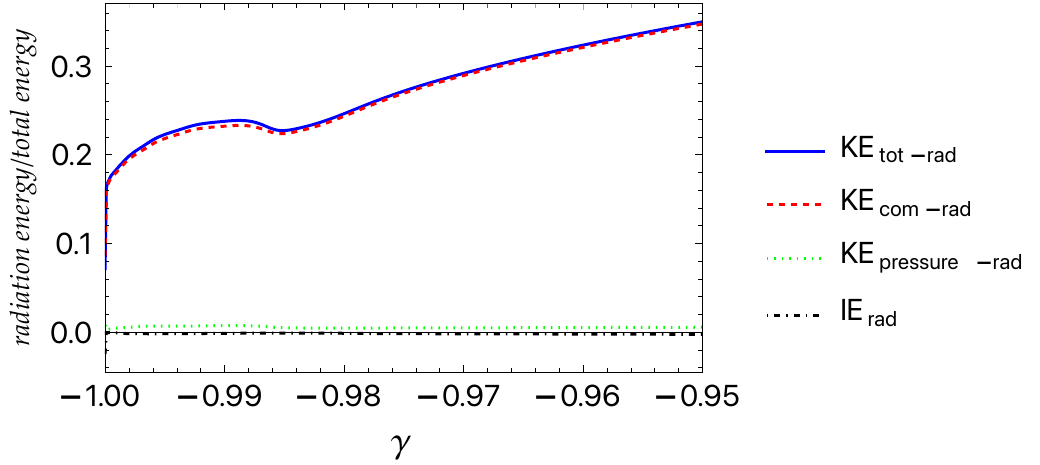}
\caption{ Energy  components of emitted radiation. $KE_{tot-rad}$: total radiation kinetic energy, $KE_{com-rad}$: center-of-mass radiation kinetic energy, $KE_{pressure-rad}$: radiation kinetic energy pressure, $IE_{rad}$: radiation interaction energy. Parameters used: $g_1=1/2,\,g_2=1,\,u_0=1, v_0=-0.64$. } 
\label{fig17}
\end{figure}

\begin{figure}[!h]    
\includegraphics[width=8.5cm]{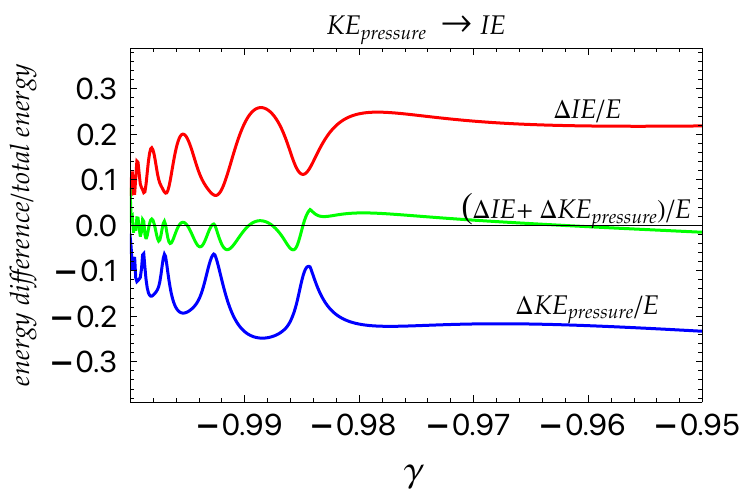}
\includegraphics[width=8.5cm]{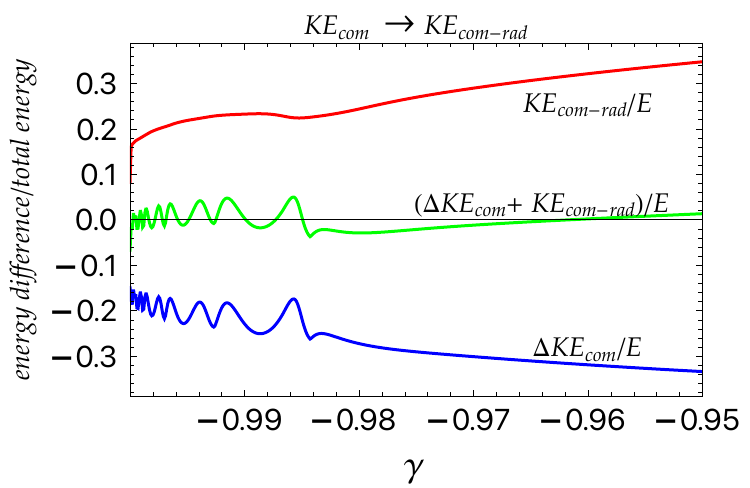}
\caption{ Change in the kinetic and interaction energies of the flat-top solitons and radiation as a results of solitons collision.  Left: mainly transfer  from solitons' kinetic energy pressure to solitons' interaction energy. Right: mainly transfer from center-of-mass kinetic energy of solitons to center-of-mass kinetic energy of radiation. Parameters used: $g_1=1/2,\,g_2=1,\,u_0=1, v_0=-0.64$.} 
\label{fig18}
\end{figure}

It is instructive to see how the above-described picture changes in terms of the width of the scattering solitons. To find out, we repeat the detailed analysis of energy transfer for a range of $\gamma$ values. We consider the range $\gamma\in[-0.95,10^{-13}-1]$. For this range, the coalesced soliton is located almost at the same position at the final time of evolution. This enables calculating the energy components of the soliton and radiation separately using  fixed integration limits. In Fig.~\ref{fig17}, we show that the total kinetic energy of radiation is mainly of the center-of-mass type. The radiation and interaction energies are very small compared to the center-of-mass or total energy. In Fig.~\ref{fig18}, we plot the difference between the final energy components of solitons  and radiation relative to  the total energy. The first subfigure shows  that the kinetic energy pressure of the initial solitons is mainly transferred to interaction energy. The second subfigure shows that the reduction in the center-of-mass kinetic energy of the initial solitons transfers mainly into the form of radiation center-of-mass kinetic energy. The latter energy transfer is the main cause of inelasticity in flat-top soliton collisions.

\section{Variational Calculation}
 \label{analsec}
In this section, we perform a variational calculation that explains the numerical results. In particular, it provides an analytical account of the amount of radiation production, and hence reveals the physics underlying the degree of analyticity of collisions between flat-top solitons. 

The main goal of the variational calculation in this section is to provide an analytical account for the amount of radiation in terms of the parameters of the system. Specifically, we want to confirm that the amount of radiation increases as the width of the flat-top soliton increases, i.e. when $\gamma$ approaches $-1$. This should explain the above numerical results which showed that completely elastic collisions for $\gamma=0$ (bright solitons), start to increasingly loose their elasticity as $\gamma$ approaches $-1$.

The variational calculation, in this section, is divided into two steps. At first, we choose an ansatz function and perform a variational calculation for a single flat-top soliton to calculate the equilibrium ground state energy and equilibrium width of the soliton. The purpose of this step is to ensure that the selected ansatz function leads to the correct static properties as compared with those obtained from the exact solution. In the second step, we use the same ansatz function for the soliton in the presence of an unknown amount of radiation background. The amplitude of the background radiation is treated as a variational parameter, which is then determined by minimizing  the energy with respect to the amount of radiation.

\subsection{Single Flat-Top Soliton}
\subsubsection{Static Case}
Choosing the appropriate ansatz function representing the profile of flat-top soliton  is a delicate step since the trial function needs to account for the `flatness' of the top of the soliton through an additional variational parameter. One needs also to  compromise between  the accuracy of the variational results and the difficulty of calculating the relevant integrations. In the literature, supergaussian is typically used  \cite{supgauss1, supgauss2}   
\begin{equation}
\psi_{\rm sup. gauss.}(x,t)=A\,e^{-\left(\frac{x-x_0}{w}\right)^s}\,e^{i\left(v(x-x_0)+\beta\,(x-x_0)^2+\phi\right)}
\label{supgauss},
\end{equation}
where $A,w,v,x_0,\beta, \phi$, are variational parameters corresponding to amplitude, width, velocity, position, chirp, and phase, respectively. The variational parameter, $s$, is specific for flat-top solitons and controls its degree of flatness. In case one is interested in only the static properties, the variational parameters  will be independent of time, while in case the dynamics of the soliton is to be obtained through the equations of motion, the variational parameters will be time-dependent. 

One may consider  other forms of the ansatz function, such as the following one constructed from $\rm tanh$ functions
\begin{equation}
\psi_{\rm tanh}(x,t)=\frac{1}{2\sqrt{c+1}}\left\{{\rm tanh}\left[d(x+w))-{\rm tanh}(d(x-w)\right]\right\}
\label{ansatztanh},
\end{equation}
where $d=\sqrt{u_0/g_1}$ and $w$ is the variational width.  Here, $u_0$ is determined by  the norm, $N$, as given by Eq.~(\ref{norm}). Another ansatz can be written in terms of piece-wise function
\begin{equation}
\psi_{\rm piece-wise}(x,t)=\left\{\begin{array}{cc}\frac{1}{\sqrt{1+c}},&|x-x_0|\le w,\\
0,&|x-x_0|>w.\end{array}\right.
\label{ansatzpiecewise}
\end{equation}

While the above three ansatz functions result in accurate estimates of the equilibrium energy and width, employing the exact solution with a variational width and amplitude, provides another ansatz that leads to the exact width and energy. This ansatz takes the form of the exact solution, (\ref{ftsol}), but with width, $w$, and amplitude, $A$, added as variational parameters
\begin{eqnarray}
	\label{ftsolvar}
\psi\left(x,t\right) &=&A\,f
\frac{1}{\sqrt{c+\mathrm{cosh}^2\left[\frac{d}{w}(x-x_{0}-v_{0}t)\right]}}\nonumber\\&\times&  e^{i \left[u_{0}t + {v_{0}\left(2(x - x_{0}) - v_{0}t\right)}/({4 g_{1}})\right]}.
\end{eqnarray}
where $f= \sqrt{\frac{2u_{0}}{g_{2}\sqrt{1+\gamma}}}$ and
\begin{equation}
u_0={g_2^2N^2\gamma}/\left({64g_1{\rm tan}^{-1}\left[\frac{\sqrt{1+\gamma}-1}{\sqrt{\gamma}}\right]^2}\right)\label{u0eq}.
\end{equation} 
For $A=w=1$, this function becomes the exact  solution (\ref{ftsol}), which is normalized to $N$. Requesting the ansatz function to be normalized to $N$ for general $A$ and $w$, we get
\begin{equation}
A=\frac{1}{\sqrt{w}}.
\end{equation}
The energy functional, (\ref{efunc}), is then calculated using this ansatz to give
\begin{equation}
E=I_x\frac{g_1}{w^2}-I_4\frac{g_2}{2w}-I_6\frac{g_3}{3w^2},
\end{equation}
where
\begin{equation}
I_x=\frac{1}{d}\int_{-\infty}^{\infty}\left(\frac{f\,d\,\cosh(x)\sinh(x)}{\left(c+\cosh^2(x)\right)^{3/2}}\right)^2dx,
\end{equation}
\begin{equation}
I_4=\frac{1}{d}\int_{-\infty}^{\infty}\frac{f^4}{\left(c+\cosh^2(x)\right)^{2}}dx,
\end{equation}
\begin{equation}
I_6=\frac{1}{d}\int_{-\infty}^{\infty}\frac{f^6}{\left(c+\cosh^2(x)\right)^{3}}dx.
\end{equation}
Minimizing the energy with respect to $w$, the condition $d E/dw=0$ gives
\begin{equation}
w_{var}=\frac{4\left(g_1 I_x-g_3I_6/3\right)}{g_2\,I_4}\label{wvar}.
\end{equation}
The width of  the soliton at half maximum is obtained  from the condition: $\psi(W)=\psi(0)/2$, and this will be given by 
\begin{eqnarray}
W&=&w_{var}\frac{\cosh^{-1}(\sqrt{4+3c})}{d}\nonumber\\
&=&2\sqrt{\frac{g_1}{u_0}}\ln\left[4+\frac{3}{\sqrt{1+\gamma}} \right.\nonumber\\& +&\left.\frac{1}{2}\,\sqrt{-4+\left(8+\frac{6}{\sqrt{1+\gamma}}\right)^2}\right],
\end{eqnarray}
which equals the width obtained from the exact solution, (\ref{width}). This confirms the  accuracy of the ansatz function (\ref{ftsolvar}), as a variational function to the flat-top soliton.

\subsubsection{Dynamic Case}
For the dynamic case, we use the ansatz
\begin{eqnarray}
	\label{ftsolvar2}
\psi\left(x,t\right) &=&
\frac{A(t)\,f}{\sqrt{c+\mathrm{cosh}^2\left[\frac{d}{w(t)}(x-x_{0}(t)-v_{0}(t)t)\right]}}\nonumber\\&\times&  e^{i \left[u_{0}t + {v_{0}(t)\left(2(x - x_{0}(t)) - v_{0}(t)t\right)+\beta(t)(x-x_0(t))^2}/({4 g_{1}})\right]}.\nonumber\\
\end{eqnarray}
we have added chirp variational parameter, $\beta(t)$, in the phase to account for the width variations with time. The lagrangian 
\begin{equation}
L=\int_{-\infty}^{\infty}\frac{i}{2}\left(\frac{\partial\psi}{\partial t}\,\psi^*-\frac{\partial\psi^*}{\partial t}\,\psi\right)\,dx-E
\end{equation}
is then calculated from which the Euler-Lagrange equations give the of equations of motion for the variational parameters:
\begin{equation}
\frac{d}{dt}\frac{\partial L}{\partial {\dot x}_0}-\frac{dL}{dx_0}=0\,\rightarrow\frac{d}{dt}v_0(t)=0,
\end{equation}
\begin{equation}
\frac{d}{dt}\frac{\partial L}{\partial {\dot v}_0}-\frac{dL}{dv_0}=0\,\rightarrow\frac{d}{dt}x_0(t)=v_0(t),
\end{equation}
\begin{equation}
\frac{d}{dt}\frac{\partial L}{\partial {\dot w}}-\frac{dL}{dw}=0\rightarrow{\dot \beta}(t)= f_3(w,\beta,\gamma),
 \end{equation}
 \begin{equation}
\frac{d}{dt}\frac{\partial L}{\partial {\dot \beta}}-\frac{dL}{d\beta}=0\rightarrow {\dot w}(t)= f_4(w,\beta,\gamma),
 \end{equation}
 where $f_3(w,\beta,\gamma)$ and $f_4(w,\beta,\gamma)$  are rather lengthy expressions and hence not shown here. The equilibrium value of $w$ is obtained from the last  equation for $w '(t)= 0$, which is identical with the  result of the static case, (\ref{wvar}). Clearly, the first two equations of motion correspond to uniform center-of-mass motion. The last two equations of motion give the dynamics of width and amplitude ($A=1/\sqrt{w}$). We have verified, by solving the equations of motion numerically, that with an initial width not equal to the equilibrium width, the soliton acquires breathing oscillations while with initial width that equals the equilibrium width, the soliton's width remains constant during the time evolution.

 \subsection{Radiation Production}
 Having established confidence in the variational ansatz function, we use it in this section to calculate the amount of radiation produced when two flat-top solitons scatter. The setup is the same as above, namely two equal solitons; one is stationary  at $x=-x_0$ and another starting motion from $x=x_0$ towards the first soliton with initial speed $v_0$. We consider the specific case when  the two solitons coalesce and move with a speed $v_c$. An amount of radiation is produced after scattering which carries some norm, energy, and momentum. The main goal here is to calculate the amount (norm) of radiation in terms of $\gamma$ that will confirm and explain the numerical results.

The calculation is outlined as follows. We calculate the energy, norm, and momentum of the  two solitons before collision. Then we calculate the same for the coalesced solitons after collision together with radiation. Imposing the conservation of energy, norm, and momentum leads to the optimum amount of radiation produced  at a specific value of $\gamma$. Good agreement with numerical results  are then obtained.

Before collision, we use the ansatz function (\ref{ftsolvar}) to calculate the energy, which turns out  to be
\begin{equation}
E_{b}=KE_{\rm com}+2KE_{\rm pressure}+IE_1+IE_2\label{eb},
\end{equation} 
where
\begin{equation}
KE_{\rm com}=g_1Nv_0^2,
\end{equation}
\begin{eqnarray}
KE_{\rm pressure}&=&\frac{g_1A^2df^2}{4(c(1+c))^{3/2}w}\nonumber\\&\times&\left(\sqrt{c(1+c)}\,(1+2c)-\sinh^{-1}\left(\sqrt{c}\right)\right),\nonumber\\
\end{eqnarray}
\begin{eqnarray}
IE_{1}&=&\frac{g_2A^4f^4w}{2(c(1+c))^{3/2}d}\nonumber\\&\times&\left(\sqrt{c(1+c)}\,\sinh^{-1}\sqrt{c}-2c\tanh^{-1}\left(\sqrt{\frac{c}{1+c}}\right)\right),\nonumber\\
\end{eqnarray}
\begin{eqnarray}
IE_{2}&=&-\frac{g_3A^6f^6w}{12(c(1+c))^{5/2}d}\left[-3\sqrt{c(1+c)}\,(1+2c)\right.\nonumber\\
&+&\left.(3+8c(1+c))\sinh^{-1}\left(\sqrt{c}\right)\right].
\end{eqnarray}
The $KE_{\rm pressure}$ term in Eq.~(\ref{eb}) is multiplied by 2 to account for two solitons and the terms $IE_1$ and $IE_2$ are the interaction energies of the cubic and quintic nonlinearities, respectively.

After collision, the two solitons coalesce and an amount of radiation is produced. The total norm of the coalesced solitons is less than the sum of the two norms of the solitons before scattering, namely
\begin{equation}
N_a=N_b-\delta,
\end{equation}
where $N_b=2N$, with $N$ being the norm of any of the two equal solitons before scattering, and $\delta$ is the total norm of radiation produced after scattering. 

The variational function after scattering is written as a superposition of a soliton part and a radiation part
\begin{equation}
\psi(x,t)=\psi_{\rm soliton}+\psi_{\rm rad}\label{psitot}.
\end{equation}
The soliton part, $\psi_{\rm soliton}$, takes the same form as that  of a single flat-top soliton, (\ref{ftsolvar}), but with  a norm equals to $N_c=2N-\delta$  and consequently a different value of $\gamma$ denoted as $\gamma_c$
\begin{eqnarray}
	\label{ftsola}
\psi_{\rm soliton}\left(x,t\right) &=& \sqrt{\frac{2u_{c}}{g_{2}\sqrt{1+\gamma_c}}} \nonumber\\&\times&
\frac{1}{\sqrt{\frac{1-\sqrt{1+\gamma_c}}{2\sqrt{1+\gamma_c}}+\mathrm{cosh}^2\left[\sqrt{\frac{u_{0}}{g_{1}}}(x-x_{0}-v_{c}t)\right]}}\nonumber\\&\times&  e^{i \left[u_{c}t + {v_{c}\left(2x  - v_{c}t\right)}/({4 g_{1}})\right]},
\end{eqnarray}
where
\begin{equation}
u_c={g_2^2N_c^2\gamma}/\left({64g_1{\rm tan}^{-1}\left[\frac{\sqrt{1+\gamma_c}-1}{\sqrt{\gamma_c}}\right]^2}\right)\label{u0eqc}
\end{equation} 
and
\begin{equation}
\gamma_c=\frac{4\tan^2\left[N_c\sqrt{g_3/g_1}/(2\sqrt{3})\right]}
{\left(\tan^2\left[N_a\sqrt{g_3/g_1}/(2\sqrt{3}\right]-1\right)^2},
\end{equation}
which are obtained from Eq.~(\ref{u0eq}) after replacing $\gamma$ by $\gamma_c$ and $N$ by $N_c$ and then solving for $\gamma_c$.  The center-of-mass speed of the coalesced soliton, $v_c$, is determined from the conservation of momentum, as follows
\begin{equation}
Nv_0=(2N-\delta)v_c,
\end{equation}
where the left hand side is the momentum of the two solitons before collision \--- one soliton is stationary \--- and the right hand side is the momentum of the coalesced solitons. We have ignored the momentum carried by  the radiation part since it  is negligible compared to that of the solitons part.

\begin{figure}[!h]    
\includegraphics[width=8.5cm]{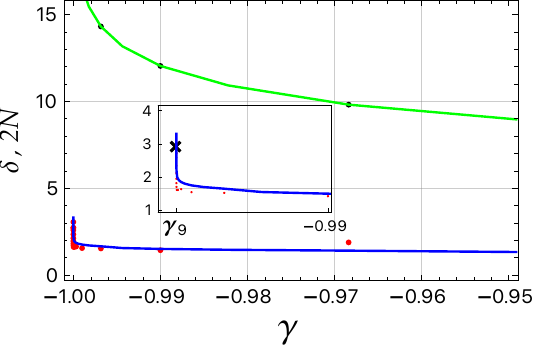}
\caption{ Lower curve: Radiation norm, $\delta$. Upper curve: total norm, $2N$. Values of $\gamma$ range from $-0.95$ to $\gamma_{10}=10^{-10}-1$. Points correspond to the numerical calculation and the curves are the results of the variational calculation. The inset shows the specific case discussed in text with the $\times$ corresponding to the value of $\delta$ found from the numerical simulations. Parameters used: $g_1=1/2,\,g_2=1,\,u_0=1/2, v_0=-0.64$. } 
\label{fignew}
\end{figure}

 The radiation part, $\psi_{\rm rad}$, is a linear dispersive wave 
\begin{equation}
\psi_{\rm rad}=A_r\,e^{i(kx-\omega t)}\label{psirad}
\end{equation} 
with wave number $k$ and frequency $\omega$, and amplitude $A_r=\sqrt{\frac{\delta}{L}}$. The parameter $L$ is the length of the spatial domain over which the radiation part  is to be integrated and gives a norm equals $\delta$, namely
\begin{equation}
\delta=\int_{-L/2}^{L/2}|\psi_{\rm rad}|^2dx.
\end{equation}
The dispersion relation for the wave number and frequency of the  radiation part  is determined from the condition that $\psi_{\rm rad}$ being a solution to the NLSE, (\ref{cqnlse}).  Inserting $\psi_{\rm rad}$ from Eq.~(\ref{psirad}) in the NLSE, (\ref{cqnlse}), this condition gives
\begin{equation}
\omega=g_2Ar^2+g_3A_r^4-g_1k^2\label{om1}.
\end{equation}
Comparing our ansatz for radiation, $\psi_{\rm  rad}$, with the exact constant-wave (CW) solution
\begin{equation}
\psi_{CW}(x,t)=\pm\sqrt{\frac{\pm g_2-\sqrt{g_2^2+4g_3u_0}}{g_3}}\,e^{i\left(u_0t+\frac{v_0}{4g_1}(2x-v_0t)\right)}
\end{equation}
we obtain $\omega=u_0-v_0^2/(4g_1)$, which upon using in (\ref{om1}) leads to
\begin{equation}
k=\frac{1}{g_1}\sqrt{A_r^2g_2+A_r^4g_3-u_0+\frac{1}{4g_1}v_0^2}.
\end{equation}
The energy of the coalesced soliton in the presence of radiation should be calculated using the ansatz function (\ref{psitot}). Since the contribution from the overlap between the soliton part and the radiation part is much smaller than that of the soliton or radiation, we ignore all terms corresponding to the overlap and thus the total energy after scattering will be approximated by that of the coalesced soliton plus that of the radiation.
Finally, equating the energy of the two flat-top solitons before collision, $E_b$, with the energy of the coalesced soliton and radiation, and solving for $\delta$, we obtain the amount of radiation produced. In Fig.~\ref{fignew}, we plot the radiation norm, $\delta$, versus $\gamma$ and compare the results with the numerical solution, where good agreement is obtained. In particular, the specific case of $\gamma_9$, which  was examined above numerically, is represented in the inset of Fig.~\ref{fignew} by the point plotted with  a $\times$. The numerical solution gives a value of $\delta=2.99$, while the variational calculation estimate is $\delta=3.1$, which is in good agreement with the numerical value.
The calculation shows also that indeed, when $\gamma$ approaches $-1$, i.e., the soliton gets deeper into the flat-top regime, the amount of radiation increases which increases the inelasticity of flat-top solitons, as we found in the numerical simulations.

\section{Conclusions and Outlook}
\label{conc}
There are three main findings in this paper: 1) While collisions of flat-top solitons are in general inelastic, a regime in the solitons' parameters space exists such that nearly elastic collisions occur. Although this regime is characterized by minimal width oscillations and radiation production, a remarkable exchange in the widths of the solitons takes place. Such a case is shown in Fig.~\ref{fig4}(b) and (f). This behavior may find an application in optical data processing as data may be stored in the width of the solitons.  2).  There is a periodic occurrence of completely elastic collision between flat-top solitons at specific values of te initial solitons' parameters including their relative phase, relative speed, and separation, as shown in Figs.~\ref{fig13}-\ref{fig15}. This periodic occurrence takes place  at values of the parameters such that the two solitons are out-of-phase, i.e. their relative phase is $\pi$, just before their collision.  In this particular case, no  width exchange takes place, no width oscillations occur, and no radiation is produced.  3) Origin of inelasticity lies in the production of radiation during the scattering process; part of the solitons' initial center-of-mass kinetic energy will be irreversibly carried away by the dispersive radiation. This is verified by Figs.~\ref{fig16}-\ref{fig18}, where the different components of  the solitons and radiation energies separately are plotted. We show that most of the radiation energy is in the form of translational kinetic energy, and much smaller part of it is in the form of kinetic energy pressure and interaction energy. 

It is observed that coalesced solitons acquire breathing-like oscillations. These oscillations can be understood within the context of Ref.~\cite{usnew}. In this work, the potential of interaction between two flat-top solitons is calculated and shown to be molecular-type in terms of the separation between solitons, i.e., there is a nonzero `bond length' where the energy is minimum. Any initial state of two solitons with separation slightly deviated from the equilibrium bond length undergoes bond oscillations. Due to the considerable overlap between the solitons, these bond oscillations resemble breathing oscillations of a single soliton. The frequency of the oscillation can be calculated analytically by linearizing in small-amplitude oscillations around the minimum of the potential. The frequency turns out to be function of $\gamma$ and the norm of the soliton, $\delta$.
In the present work, the oscillations observed in our simulations after the two colliding solitons coalesce can be viewed as those of the two-soliton molecule. From the variational calculation in Sec.~\ref{analsec}, the norm of the coalesced soliton is expressed as the total norm of the two initial solitons, $2N$, minus the  norm of radiation produced, $\delta$. In addition,  $\gamma$ is also different than that of the initial solitons. The frequency of the coalesced soliton, is thus a function of the norm of radiation, $\delta$ and $\gamma$.

The main questions posed at the introduction of this paper, are thus answered. Collisions of flat-top solitons can be elastic as long as the relative phase between the two colliding solitons equals $\pi$. In such a case, the solitons do not overlap at the closest approach as a result of the repulsive force. Consequently, no production of radiation occurs and hence total kinetic energy will be conserved indicating an elastic scattering. Relative phase at collision can be monitored by the initial parameters of the solitons, namely their initial relative phase, initial relative speed, and initial separation. When any of these initial parameters change by an amount leading to a change in the relative phase that is a multiple of $2\pi$, the same scattering outcome will be obtained, which explains the periodic appearance of elastic scattering. 

Inelasticity is directly caused by the production of radiation, and production of radiation is caused by the nonintegrability of the NLSE equation with dual power law nonlinearity, which is confirmed here.

On the fundamental level, we believe the present work lays down the essentials of two theoretical investigations. Firstly, investigating the link between nonintegrability and inelasticity through the calculation of conserved quantities. One needs to find out which physical quantity that is not conserved and is responsible for inelasticity. Our preliminary investigations shows that this quantity must  be other than norm, energy, or momentum, as these three quantities turn out to be conserved for the NLSE with dual nonlinearities. The second investigation is a variational calculation that accounts for the intensity of radiation and loss in the solitons kinetic energy. In view of the above findings, specifically Figs.~\ref{fig17} and \ref{fig18}, the trial function may be set up with a soliton part and a radiation part. The variational parameters may be the amplitudes and speeds of the soliton and radiation. These two theoretical works are left for future investigations.

Finally, the results we found here may be useful for applications of flat-top solitons such as data carriers, where elastic collisions are favored due to minimization of errors  \cite{app0,app1,app2}. Specifically, optical data transfer and data processing favor elastic scattering in order to minimize errors. Data may be coded in the amplitude or width of the soliton. Regimes of elastic scattering identified in this paper would be useful in this case. In addition, the interesting  regimes of  nearly elastic scattering but with considerable exchange in width (the `nearly elastic scattering' regime in Fig.~\ref{fig6})  would be particularly important. Data may be coded in the ratio between the  widths of the solitons. Scattering will change this ratio. The change depends on several parameter including initial separation, relative velocity, relative phase, and gamma. Manipulating  these parameters a certain protocol may be designed  to obtain the function of operations such as logic gates.

\end{document}